\documentclass[11pt]{article} 
\usepackage{mystyle-new}
\usepackage{authblk}
\usepackage[T1]{fontenc}
\usepackage{epsfig,amsmath} 
\usepackage{hyperref}
\usepackage{color}
\usepackage{graphicx}
\usepackage{orcidlink}
\definecolor{red}{rgb}{1,0,0}
\def\lesssim{\ \hbox{\raise 2pt \hbox{$<$} \kern -13pt
                     \lower 3pt \hbox{$\sim$}}\ }
\def\greatersim{\ \hbox{\raise 2pt \hbox{$>$} \kern -13pt
                     \lower 3pt \hbox{$\sim$}}\ }

\def\lsim{\mathrel{\rlap{\lower4pt\hbox{\hskip1pt$\sim$}}
    \raise1pt\hbox{$<$}}}                
\def\gsim{\mathrel{\rlap{\lower4pt\hbox{\hskip1pt$\sim$}}
    \raise1pt\hbox{$>$}}}                

\input epsf.tex
\def\desepsf(#1 width #2){\epsfxsize=#2 \epsfbox{#1}}
\def\kt{\ensuremath{k_{\rm T}}}

\def\qt{\ensuremath{q_{\rm t}}}
\def\qtmin{\ensuremath{q_{min}}}

\def\zM{\ensuremath{z_{\rm M}}}

\newcommand{\alphas}{\ensuremath{\alpha_\mathrm{s}}}
\newcommand{\alphaem}{\ensuremath{\alpha_\mathrm{em}}}

\newcommand{\as}{\ensuremath{\alpha_\mathrm{s}}}

\newcommand{\PB}{PB}
\newcommand{\PBset}{{PB-NLO-2018}}
\newcommand{\PBnewQCD}{{PB-NLO-QCD-2025}}
\newcommand{\PBnewEW}{{PB-NLO-QCD+EW-2025}}

\newcommand{\GeV}{\text{GeV}}

\def\PZ{\ensuremath{{\rm Z}}}
\def\PW{\ensuremath{{\rm W}}}

\def\Pgamma{\ensuremath{\gamma}}
\def\Pgluon{\ensuremath{\rm g}}

\def\swtwo{\sin^2 \theta_\PW }
\def\cwtwo{\cos^2 \theta_\PW }
\def\swfour{\sin^4 \theta_\PW }
\def\cwfour{\cos^4 \theta_\PW }

\newcommand{\alphaeff}{\ensuremath{\alpha_\mathrm{eff}}}

\def\updfevolv{{\sc uPDFevolv2}}

\newcommand{\nnloSplit}{vanNeerven:2000wp,Moch:2004pa,Vogt:2004mw,Vermaseren:2005qc,Blumlein:2021enk,Blumlein:2022gpp,ABLINGER2014263,Ablinger:2017tan,Moch:2014sna,Behring:2019tus,Blumlein:2021ryt}

\newenvironment{tolerant}[1]{\par\tolerance=#1\relax}{ \par }
\usepackage{amsmath,bm}
\usepackage{lineno}

\usepackage{cite,./mcite}
\usepackage{tikz}
\usepackage[symbol]{footmisc}

\newcommand{\dglap}{Gribov:1972ri,Lipatov:1974qm,Altarelli:1977zs,Dokshitzer:1977sg}

\providecommand{\DOI}[1]{\href{http://dx.doi.org/#1}}

\begin{document}

\title{Determination of Parton Densities for QCD partons and Electroweak Bosons} 
\author[1]{K.~Moral~Figueroa\orcidlink{0000-0003-1987-1554}}
\author[1]{E.~Gallo\orcidlink{0000-0001-7200-5175}}
\author[1,2,3]{H.~Jung\orcidlink{0000-0002-2964-9845}}
\affil[1]{Deutsches Elektronen-Synchrotron DESY, Germany}
\affil[2]{Elementary Particle Physics, University of Antwerp, Belgium}
\affil[3]{II. Institut f\"ur Theoretische Physik, Universit\"at Hamburg,  Germany}
\author[1]{S.~Taheri~Monfared\orcidlink{0000-0003-2988-7859}}

\date{}
\begin{titlepage} 
\maketitle
\vspace*{-8cm}
\begin{flushright}
DESY-25-150\\
\end{flushright}
\vspace*{+10cm}

\begin{abstract}
Parton densities are obtained from a solution of the extended DGLAP-type evolution equation that includes both QCD and electroweak contributions. The equations are solved using the Parton-Branching (\PB ) approach, and the evolution is performed at next-to-leading order for QCD partons and leading order for electroweak bosons.

The initial  QCD parton distributions are fitted to HERA deep inelastic scattering data, while photon and weak-boson densities are generated perturbatively and validated against  $d\sigma/dQ^2$.

The resulting collinear and transverse-momentum dependent (TMD) densities are provided in LHApdf and TMDlib formats for direct phenomenological use.
\end{abstract} 
\end{titlepage}

\section {Introduction}
Predictions for high-energy hadronic processes rely on the factorization of perturbative and non-perturbative components.
The parton densities themselves consist of a perturbative part (the perturbative evolution equations) and the non-perturbative input distributions at a scale of the order of the hadron mass. 
Several collaborations, such as CTEQ~\cite{CTEQ-page,Hou:2019efy,Xie:2023qbn}, MSHT~\cite{MSHT-pape,Harland-Lang:2019pla,Cridge:2021pxm,Cridge:2023ryv} and NNPDF~\cite{NNPDF-page,NNPDF:2024nan} have performed precise global fits of QCD parton densities. A public tool, xFitter~\cite{xFitterDevelopersTeam:2022koz,Alekhin:2014irh},  exists for the fitting framework. QCD parton and photon densities are now determined by all groups, but heavy boson densities are not yet  available in parameterized form (in the parton-density repository LHAPDF~\cite{Buckley:2014ana}).
In this publication we describe the determination of a complete set of parton densities\footnote{We use the wording {\it parton density} for partons but also for electroweak bosons.} for QCD partons as well as for the electroweak bosons: \Pgamma , \PZ\ and \PW .

\begin{tolerant}{9000}
The factorization framework simplifies complex hadronic processes and offers physical insight through the picture of parton branching.
The Effective W/Z Approximation (EWA) ~\cite{Dawson:1984gx,Kane:1984bb,Kunszt:1987tk,Chanowitz:1985hj} involves the partitioning of collinear, initial-state \PW /\PZ\ boson emissions out of matrix elements (MEs)  into collinear parton density functions (PDFs). This approximation, as an extension of the Equivalent Photon Approximation (EPA)~\cite{Budnev:1975poe,Frixione:1993yw} (or  Weizs\"acker-Williams Approximation~\cite{vonWeizsacker:1934nji,Williams:1934ad}), offers significant computational benefits, particularly dealing with infrared limits of phase space and avoiding numerical instabilities in  calculations when scale hierarchies are present. 
 A  concise derivation of the equivalent vector-boson approximation is given in Ref.~\cite{Alikhanov:2018kpa}.
The concept of EWA has already been used in calculations including weak vector boson fusion (VBF) \cite{Brehmer:2014pka,Hagiwara:2009wt,Dicus:1987ez,Altarelli:1987ue,Duncan:1985vj} and heavy quark production via  \PW \Pgluon -scattering \cite{Willenbrock:1986cr,Brehmer:2014pka}. 
Renewed interest in the EWA came with the discussion of high energy muon colliders \cite{Ruiz:2021tdt}.
 \end{tolerant}

Recently, collinear parton densities including electroweak (EW) bosons have been  discussed in more complete approaches~\cite{Fornal:2018znf,Bauer:2017bnh,Bauer:2017isx,Chen:2016wkt}. Electroweak  evolution equations are discussed in \cite{Dittmaier:2025htf,Ciafaloni:2001mu,Ciafaloni:2005fm}. Splitting functions for \PW -emission including a transverse momentum dependence have been calculated in Ref.\cite{Bagdatova:2024aem}. 
On the experimental side, measurements of jets and EW bosons at the LHC\cite{Mikel-PHD-2023,ATLAS-CONF-2021-033} support the picture, that at high energies the radiation of massive EW bosons from light quarks \cite{Christiansen:2014kba} can be treated similarly to those of massless photons or gluons. 

In this study, we present a solution of the extended DGLAP evolution equation including the photon and heavy electroweak bosons. We apply the Parton Branching (\PB ) method~\cite{Hautmann:2017fcj,Hautmann:2017xtx} for solving the evolution equation. 
This method (as already implemented in the evolution package \updfevolv~\cite{Jung:2024uwc}) has the advantage that implementing additional boson contributions and kinematic constraints is straightforward.
Together with the determination of collinear parton densities, Transverse Momentum Dependent (TMD) densities are also calculated. 

We discuss in Chapter~\ref{equations} the evolution equations and introduce the splitting functions and couplings for EW bosons. We also discuss the treatment of heavy boson masses.  In Chapter~\ref{QCDfits} we describe the fit of the initial conditions of the strongly interacting partons to precision measurements obtained at HERA and  the determination of parton  distributions.
We also show a validation of the heavy boson densities, using measurements of $d\sigma/dQ^2$ obtained at HERA for both neutral and charged currents. Conclusions are given in Chapter~\ref{Conclusion}.

\section{Evolution equations for QCD partons and electroweak bosons\label{equations}}

The evolution of parton densities with scale $\mu^2$ is described by the DGLAP equation~\cite{\dglap}:
\begin{equation}
\label{EvolEq}
 \mu^2 \frac{{\partial }{x f}_a(x,\mu^2)}{{\partial } \mu^2}   =  
 \sum_b
\int_x^{1} dz \; {P}_{ab} \left(\alphaeff(\mu^{2}),z\right)  \; \frac{x}{z}{f}_b\left({\frac{x}{z}},
\mu^{2}\right)   \; ,
\end{equation}
where  $P_{ab}$ denotes the regularized splitting function for the transition  $b \to a$, which can be decomposed as (in the notation of Ref.~\cite{Hautmann:2017fcj}):
\begin{equation}
{P}_{ab}(z,  \alphaeff) = D_{ab}(  \alphaeff)\delta(1-z) + K_{ab}( \alphaeff)\frac{1}{(1-z)_{+}} + R_{ab}(z, \alphaeff)  \; .
\label{Eq:Pdecomp}
\end{equation}

The splitting functions for QCD partons are known, and summarized for NLO and NNLO in Ref~\cite{\nnloSplit}. Photon and heavy-boson radiation from quarks occurs analogously to gluon emission. In the PB approach, the evolution equations are rewritten using Sudakov form factors, which replace the plus-prescription and allow a probabilistic interpretation, as discussed in detail in Refs.~\cite{Hautmann:2017fcj,Hautmann:2017xtx}.
The coupling \alphaeff\ is the effective coupling, dependent on the parton (particles) involved, for QCD partons $\alphaeff = \alphas$.

The evolution equation in terms of Sudakov form factors $\Delta^S_a( \mu^2)$ is given by: 
\begin{equation}
  {x f}_a(x,\mu^2)  =  \Delta^S_a (  \mu^2  ) \  {x f}_a(x,\mu^2_0)  
+ \sum_b
\int^{\mu^2}_{\mu^2_0} 
{{d q^2 } 
\over q^2 } 
{
{\Delta^S_a (  \mu^2  )} 
 \over 
{\Delta^S_a( q^2
 ) }
}
\int_x^{\zM} {dz} \;
 \frac{\alphaeff}{2\pi} \hat{P}_{ab} (z) \frac{x}{z}
\;{f}_b\left({\frac{x}{z}},
q^2\right)  \; .
\end{equation}
The limiting scale  $\zM $ in the $z$-integral is important in this prescription,  since  to be consistent with the DGLAP equations for massless QCD partons and the photon, the integration limit  $\zM \to 1$ (for the numerical calculations we use  $\zM = 1 - \epsilon$, with $\epsilon$ being  small). 
The Sudakov form factor is given by:
\begin{equation}
\label{sud-def}
  \Delta^S_a (\mu^2 ) = \Delta^S_a (\zM, \mu^2 , \mu^2_0 ) = 
\exp \left(  -  \sum_b  
\int^{\mu^2}_{\mu^2_0} 
{{d { q}^{ 2} } 
\over {q}^{2} } 
 \int_0^{\zM} dz \  z  \frac{\alphaeff}{2\pi}
\ P_{ba}^{(R)}\left( z \right) 
\right) 
  \;\; ,   
\end{equation}

In the \PB -approach, the evolution equation for collinear densities is extended to include also transverse momenta of the partons (and bosons). After identifying the evolution scale with a physical scale, here  the rescaled transverse momentum (coming from angular ordering),  transverse momenta of the emitted partons can be calculated, as described in detail in Ref.~\cite{Hautmann:2017fcj,Hautmann:2017xtx}.

The calculation of collinear photon densities has been discussed quite extensively in literature, see e.g. Refs.~\cite{Jung:2021mox,Manohar:2017eqh,Schmidt:2015zda,Manohar:2016nzj,Ball:2013hta,Martin:2004dh,Roth:2004aa,Gluck:2002fi,Harland-Lang:2019pla}. For a complete description of the photon density, also an intrinsic distribution of photons inside the hadron must be included, which is relevant at small scales $\mu^2$, while at large scales the dominant contribution comes from photon radiation off the quarks. Several groups~\cite{Martin:2004dh,Schmidt:2015zda,Harland-Lang:2019pla} allow and treat an intrinsic photon component inside the proton. In the approach described here, we neglect any initial photon contribution, since we are interested in the contribution at large scales and aim for a consistent treatment of photon and heavy boson densities.
The determination of effective \PW\ densities has already been  discussed  in Refs.~\cite{Kane:1984bb,Lindfors:1985yp,Cahn:1984tx,Dawson:1984gx,Chanowitz:1985hj,Kleiss:1986xp,Dawson:1986tc,Altarelli:1987ue,Kunszt:1987tk} and in recent years, these ideas have been revisited  in Refs.~\cite{Ciafaloni:2005fm,Bauer:2017isx,Bauer:2017bnh,Fornal:2018znf,Ciafaloni:2024alq}.

The effective coupling $\alphaeff$ (given in Tab.~\ref{Tab:GenSplitt}) for photons is $\alphaeff = \alphaem$, while for processes with \PZ\ or \PW\ it can be obtained from the total production $q_i {\bar q}_j$-cross section, and is given by:
\begin{eqnarray}
 \PZ: \alphaeff & = & \frac{\alphaem }{4 \swtwo \cwtwo } (V_f^2 + A_f^2)  \\ 
 \PW: \alphaeff & =  &  \frac{\alphaem |V_{qq}|^2}{4 \swtwo  } \; ,
 \end{eqnarray}
 with $\theta_W$ being the Weinberg angle and  $V_f$, $A_f$ being the  vector and axial couplings of the \PZ -boson to the fermions, and $V_{qq}$ being the CKM element.

In the massless limit, the splitting functions for EW processes at LO are identical to the corresponding ones of QCD partons modulo the coupling and  different color factors (as given in Tab.~\ref{Tab:GenSplitt}), and the self-coupling, which does not exist  for photons (the treatment of heavy boson masses is discussed in Section~\ref{bosons}).
In Appendix~\ref{EWsplitt} we give an overview of the EW splitting functions at leading order. We average over all transverse polarization states, and neglect any longitudinal contribution from \Pgamma -, \PZ - and \PW - contribution (only in section~\ref{validation} for the validation with the DIS cross section, we need to obtain \PW - densities with different polarization states).
A full account including different polarization states is given in Refs.~\cite{Ciafaloni:2024alq,Ciafaloni:2005fm,Fornal:2018znf,Chen:2016wkt,Bauer:2017isx}.

We show here explicitly the momentum weighted density for photons, \PZ - and \PW -bosons:
\begin{eqnarray}
 \mu^2 \frac{{\partial }{x f}_{\gamma}(x,\mu^2)}{{\partial } \mu^2} &  =  &
\int_x^{1} dz \; {P}_{\gamma q} \left(\alphaeff(\mu^{2}),z\right)  \;  \sum_i  e_i^2 \left[  \frac{x}{z}{f}_i\left({\frac{x}{z}}, \mu^{2}\right) +  \frac{x}{z}{\bar f}_i\left({\frac{x}{z}}, \mu^{2}\right) \right]  \nonumber \\
& = &  \frac{\alphaem}{2 \pi}  \int_x^{1} \frac{dz}{z}\left(1 + ( 1- z) ^2 \right)  \sum_{u,d}  e_i^2 \left[  \frac{x}{z}{F}_i\left({\frac{x}{z}}, \mu^{2}\right) +  \frac{x}{z}{\bar F}_i\left({\frac{x}{z}}, \mu^{2}\right) \right]  \label{gammaPDF}\\ 
 \mu^2 \frac{{\partial }{x f}_{\PZ}(x,\mu^2)}{{\partial } \mu^2} & = &  \frac{\alphaem}{8 \pi \swtwo \cwtwo}  \int_x^{1} \frac{dz}{z}\left(1 + ( 1- z) ^2 \right)  \nonumber \\ 
 & & \times \sum_{u,d}  \left(V_i^2 +A_i^2\right)  \left[  \frac{x}{z}{F}_i\left({\frac{x}{z}}, \mu^{2}\right) +  \frac{x}{z}{\bar F}_i\left({\frac{x}{z}}, \mu^{2}\right) \right]  \label{ZPDF}\\ 
 \mu^2 \frac{{\partial }{x f}_{\PW^-}(x,\mu^2)}{{\partial } \mu^2} & = &  \frac{\alphaem}{8 \pi \sin^2\theta_W}  |V_{qq}|^2}{ \int_x^{1} \frac{dz}{z}\left(1 + ( 1- z) ^2 \right)   \frac{x}{z} \left[{D}\left({\frac{x}{z}}, \mu^{2}\right) + {\bar U}\left({\frac{x}{z}}, \mu^{2}\right) \right]  \label{W-PDF}\\ 
 \mu^2 \frac{{\partial }{x f}_{\PW^+}(x,\mu^2)}{{\partial } \mu^2} & = &  \frac{\alphaem}{8 \pi \swtwo}  |V_{qq}|^2}{ \int_x^{1} \frac{dz}{z}\left(1 + ( 1- z) ^2 \right)   \frac{x}{z} \left[{U}\left({\frac{x}{z}}, \mu^{2}\right) + {\bar D}\left({\frac{x}{z}}, \mu^{2}\right) \right] \label{W+PDF} 
\end{eqnarray}
where we have introduced the following combinations of the quark densities $f_i(x,Q^2)$:
\begin{eqnarray}
xF_u=xU=xu+xc, x{\bar F}_u=x{\bar U }=x{\bar u} +x{\bar c}, \\
xF_d=xD=xd+xs, x{\bar F}_d = x{\bar D} =x{\bar d} +x{\bar s}
\end{eqnarray}

\subsection{Mass treatment for Heavy Bosons\label{bosons}}

In  the Zero-Mass-Variable-Flavor-Number-Scheme (ZMVFN) used for QCD partons,  all partons are treated  massless. Heavy-flavors are only produced, once the scale is above the heavy flavour mass. 
An analogous scheme is applied for heavy bosons. However, measurements of charged-current cross sections in lepton-proton collisions give a non-zero cross section  for $Q^2 < M^2_{\PW} $, suggesting the application of a suppression factor $\left[\frac {Q^2}{Q^2 + M_{\PW}^2}\right]^2$. In an approach applying angular ordering, the mass of the emitted (or emitting) parton can limit the range in $z$-integration by $\zM = 1 - \frac{M}{q}$.

We now summarize three mass-treatment scenarios (referred to as {\sc MassCutScheme}): 
\begin{itemize}
\item {\sc MassCutScheme=0}:  In the ZMVFN scheme, the heavy boson mass $M_V$ is included as a mass threshold as applied in the EWA~\cite{Ruiz:2021tdt}:
\begin{equation}
\alphaeff \equiv \alphaeff \cdot \Theta\left(q^2 - M_V^2\right)
\end{equation}
\item {\sc MassCutScheme=1}: The boson mass $M_V$ is  treated by a suppression factor~\cite{Fornal:2018znf} (similar to the treatment in the DIS cross section):
\begin{equation}
\alphaeff \equiv \alphaeff \cdot \left[\frac {q^2}{q^2 + M_V^2}\right]^2 \; .
\end{equation}
This is taken as the default scenario.
\item {\sc MassCutScheme=2}: In angular ordering, the boson mass $M_V$ is included via a limit in the $z$ integrals~\cite{Bauer:2017isx}  (and thus also in \alphaeff ):
\begin{equation}
\alphaeff \equiv \alphaeff \cdot  \Theta\left(q^2(1-z)^2 - M_V^2 \right)  \;\; {\rm and } \;\; \zM = 1 - \frac{M_V}{q} 
\end{equation}
\end{itemize}
In the first two cases, $\zM \to 1$, while in {\sc MassCutScheme=2}  the $z$-integration is limited. In all cases, the splitting functions are taken in the massless limit.
The different choices will lead to different collinear and in particular, transverse momentum dependent distributions.

\section{QCD fits of initial parton distributions\label{QCDfits} }

We determine the initial collinear and TMD distributions using the \PB -method, fitting to precision DIS cross-section data from HERA~\cite{Abramowicz:2015mha}, as detailed in Ref.~\cite{Martinez:2018jxt}.
The application of the extended evolution equation eq.~(\ref{EvolEq}), including photons and electroweak bosons requires a new determination of the initial parameters, because  additional radiation of photons and electroweak bosons off quarks will lead to a reduction of the quark densities. 
The determination of the parton densities is performed under the following assumptions: a.) the initial photon density is ignored in order to keep the same structure as for heavy boson densities, b.) interference of \Pgamma\ and \PZ\ is ignored during the evolution, and c.) the DIS structure functions are unchanged compared to the NLO QCD case, since the electroweak contributions are considered only at LO (see discussion in Ref.~\cite{Giuli:2017oii}).

We perform new fits of the initial parameters to obtain a complete set of QCD and EW parton densities.
We follow the same strategy as in Ref.~\cite{Martinez:2018jxt} and determine two different sets of distributions, labeled as Set1\footnote{Although, for the transverse momentum distributions of Drell-Yan (DY) lepton pair, Set2 (with $\as(\qt)$) describes the measurements much better (see Refs.~\cite{Bubanja:2023nrd,Bubanja:2024puv}), for most of the  calculations using collinear parton distributions the strategy of Set1 (using $\as(q)$) is used.}
 and Set2: in Set 1, we use $\as(q)$ while in Set 2 $\as(\qt )$ is applied with $q$ being the virtuality of the propagating parton, and \qt\ is the transverse momentum of the emitted parton with $\qt = q (1-z)$.
 
Studies of the transverse momentum distribution of DY lepton pairs at the LHC showed, that the limit of  $\zM = 1 - q_0/q$ with $q_0=0.01$ \GeV\  as used in \PBset~Set2 was leading to a small energy dependence of the determined intrinsic \kt -distribution, while using a much smaller $q_0=0.00001$ \GeV\ gave an $\sqrt{s}$ independent intrinsic \kt -distribution~\cite{Bubanja:2024puv}. For the new densities, which we label \PBnewEW , we take this observation into account by consistently using \zM\ as a technical parameter with $\zM \to 1$ using $\zM = 0.99999$.

The QCD part of the evolution is treated at next-to-leading order (NLO) in both \as\ and the splitting functions, while the electroweak part (photon and electroweak boson) is taken with leading order splitting functions. This so-called "phenomenological" approach~\cite{Manohar:2017eqh} is taken, since  $\as^2 \sim \alpha_{\rm em}$, in contrast to a "democratic" approach, where the same order in the perturbative expansion is used.
In the calculations we take $\alpha_{\rm em} = 1/137$ fixed  since different scale choices for electroweak processes are less clear and require a further detailed study. The scheme of electroweak parameters used  is given in Appendix~\ref{EWinput}. The heavy boson mass is treated with {\sc MassCutScheme=1}. The other two schemes, discussed in the previous section, are used for comparison only.

We use the xFitter package~\cite{xFitterDevelopersTeam:2022koz} and use several advantages in the new package (multi-thread, Ceres minimizer better suited for Monte Carlo generated grid files) compared to the fit reported in Ref.~\cite{Martinez:2018jxt}.
We apply the same form of parameterization for the QCD starting distributions, same starting scale and heavy flavor masses. 
Including photon and heavy-boson evolution slightly increases $\chi^2$ compared to the pure-QCD fit (Tab.~\ref{Fit_chi2}).
\begin{table}[htb]
\renewcommand*{\arraystretch}{1.0}
\centerline{
\begin{tabular}{|l|c|c|c|}
\hline
\multicolumn{4}{|c|}{PB NLO Set 1 $\alphas(q^2)$ } \\
\hline
\multicolumn{1}{|c|}{ } &
\multicolumn{1}{c|}{$\chi^2$} &
\multicolumn{1}{c|}{d.o.f} &
\multicolumn{1}{c|}{$\chi^2/$d.o.f} \\
\hline
\PBnewQCD\ Set1   & 1361   & 1131  &   1.20 \\
\hline
\PBnewEW\ Set1   & 1378   & 1131  &   1.22  \\
\hline
\end{tabular}}
\centerline{
\begin{tabular}{|l|c|c|c|}
\hline
\multicolumn{4}{|c|}{PB NLO Set 2 $\alphas(\qt^2)$ } \\
\hline
\multicolumn{1}{|c|}{ } &
\multicolumn{1}{c|}{$\chi^2$} &
\multicolumn{1}{c|}{d.o.f} &
\multicolumn{1}{c|}{$\chi^2/$d.o.f} \\
\hline
\PBnewQCD\ Set2   & 1388   & 1131  &   1.23  \\
\hline
\PBnewEW\ Set2   & 1396  &  1131   &  1.24 \\
\hline
\end{tabular}}
\caption{\small Values of $\chi^2$ for the different Sets at NLO.} 
\label{Fit_chi2}
\end{table}
\subsection{Uncertainties\label{uncertainties}}
Uncertainties of the distributions are obtained, as described in Ref.~\cite{Martinez:2018jxt}:  uncertainties coming from the statistical and systematic uncertainties of the measurements are estimated using the Pumplin method~\cite{Pumplin:2001ct} (as implemented in xFitter), and model uncertainties are obtained by new fits with changed heavy quark masses, starting values $\mu_0^2$  for the evolution and  \qtmin\ for Set2, as described in Tab.~\ref{Fit_model}.

The most dominant part of the uncertainties  of the electroweak densities comes from the uncertainties of the quark distributions at large $x$, both from the statistical and systematic uncertainties coming from the measurements as well as from model uncertainties of the QCD partons. Additional uncertainties of the electroweak contributions come from uncertainties in the coupling, as well as the masses of the heavy bosons (for the uncertainties given by PDG~\cite{ParticleDataGroup:2024cfk} see Appendix~\ref{EWinput}). These uncertainties are much smaller than those coming from the uncertainties of the determination of the QCD parton distributions. We have varied $m_\PZ$ and $m_\PW$ by $\pm 0.1~\GeV $ and calculated the distributions and found effects much smaller than the other model uncertainties, and thus neglected those variations.
\begin{table}[htb]
\renewcommand*{\arraystretch}{1.0}
\centerline{
\begin{tabular}{|c|c|c|c|}
\hline
\multicolumn{1}{|l|}{ } &
\multicolumn{1}{c|}{Central } &
\multicolumn{1}{c|}{Lower} &
\multicolumn{1}{c|}{Upper}  \\
\multicolumn{1}{|c|}{ } &
\multicolumn{1}{c|}{value} &
\multicolumn{1}{c|}{value} &
\multicolumn{1}{c|}{value}  \\
\hline
\PBnewEW\ Set 1  $\mu^2_0$ (GeV$^2$)	      & 1.9  & 1.6  & 2.2 \\
\PBnewEW\ Set 2  $\mu^2_0$ (GeV$^2$)	      & 1.4  & 1.1  & 1.7 \\
\PBnewEW\ Set 2  \qtmin  (GeV)	      & 1.0  & 0.9  & 1.1 \\
$m_c$  (GeV)          & 1.47 & 1.41 & 1.53 \\
$m_b$  (GeV)         & 4.5 & 4.25 & 4.75 \\
\hline
\end{tabular}}
\caption{\small Central values and change ranges of parameters for model dependence} 
\label{Fit_model}
\end{table}

\subsection{QCD parton densities}

\begin{figure}[h!tb]
\begin{center} 
\includegraphics[width=0.325\textwidth,angle=0]{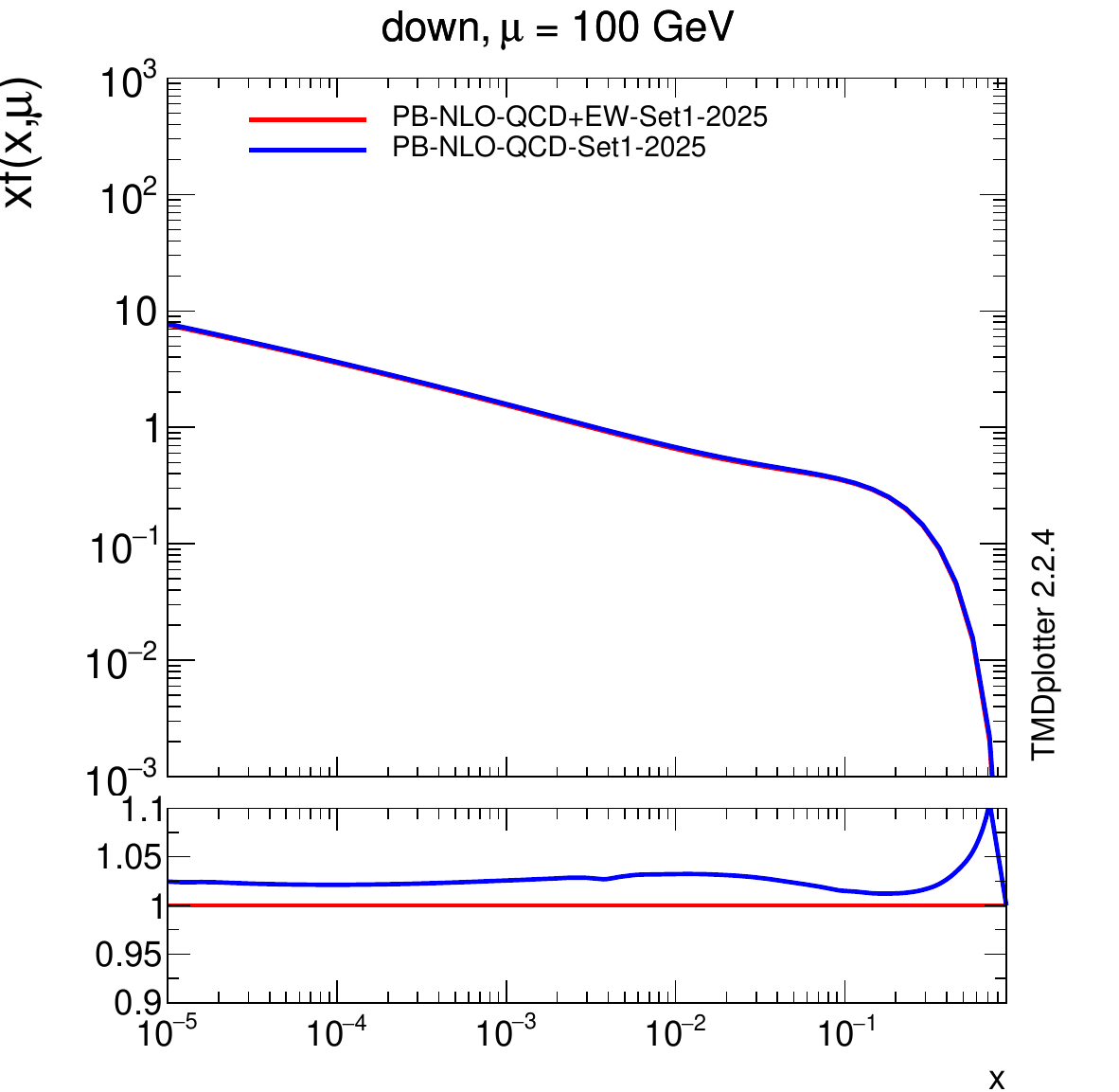}
\includegraphics[width=0.325\textwidth,angle=0]{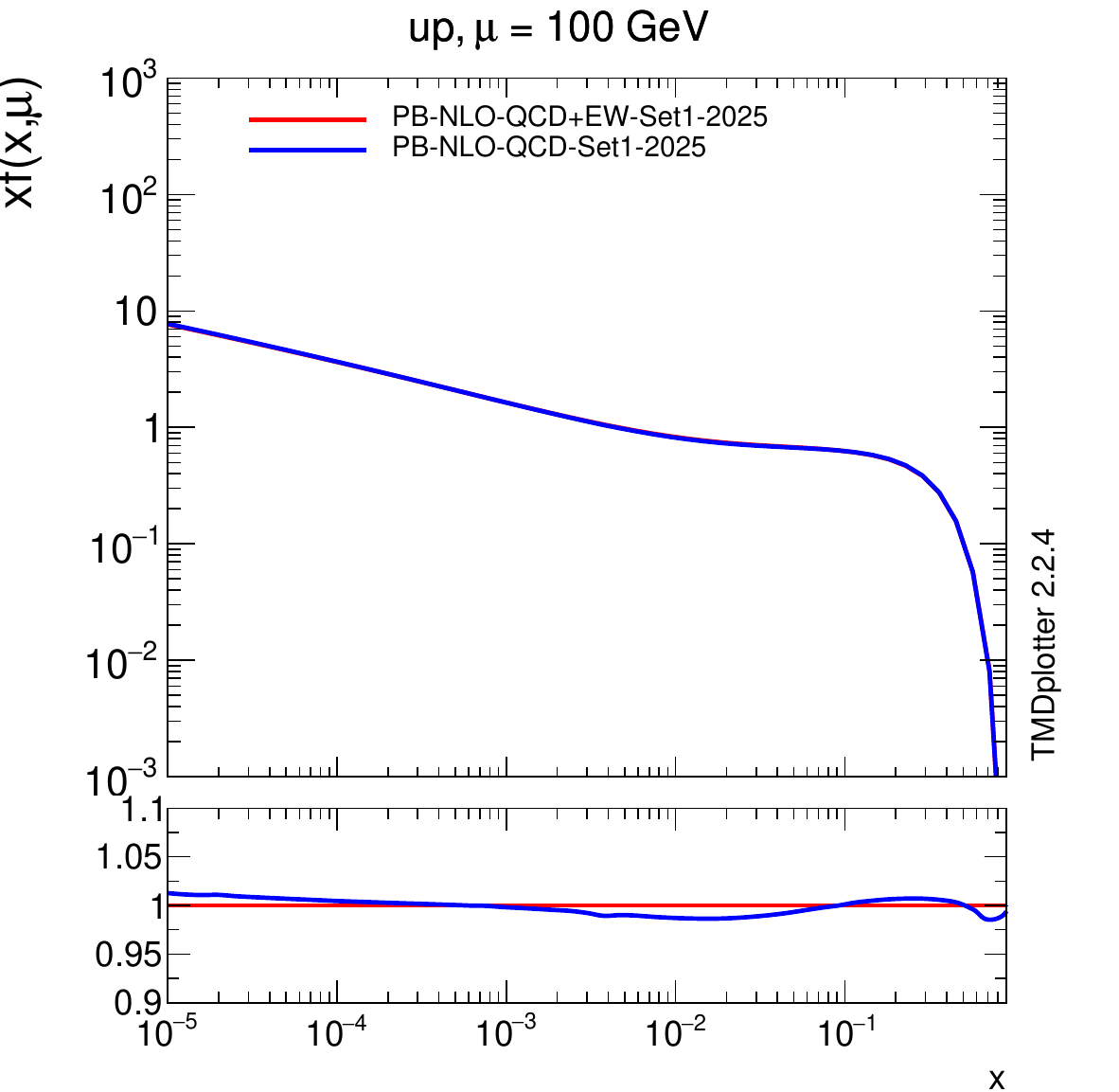}
\includegraphics[width=0.325\textwidth,angle=0]{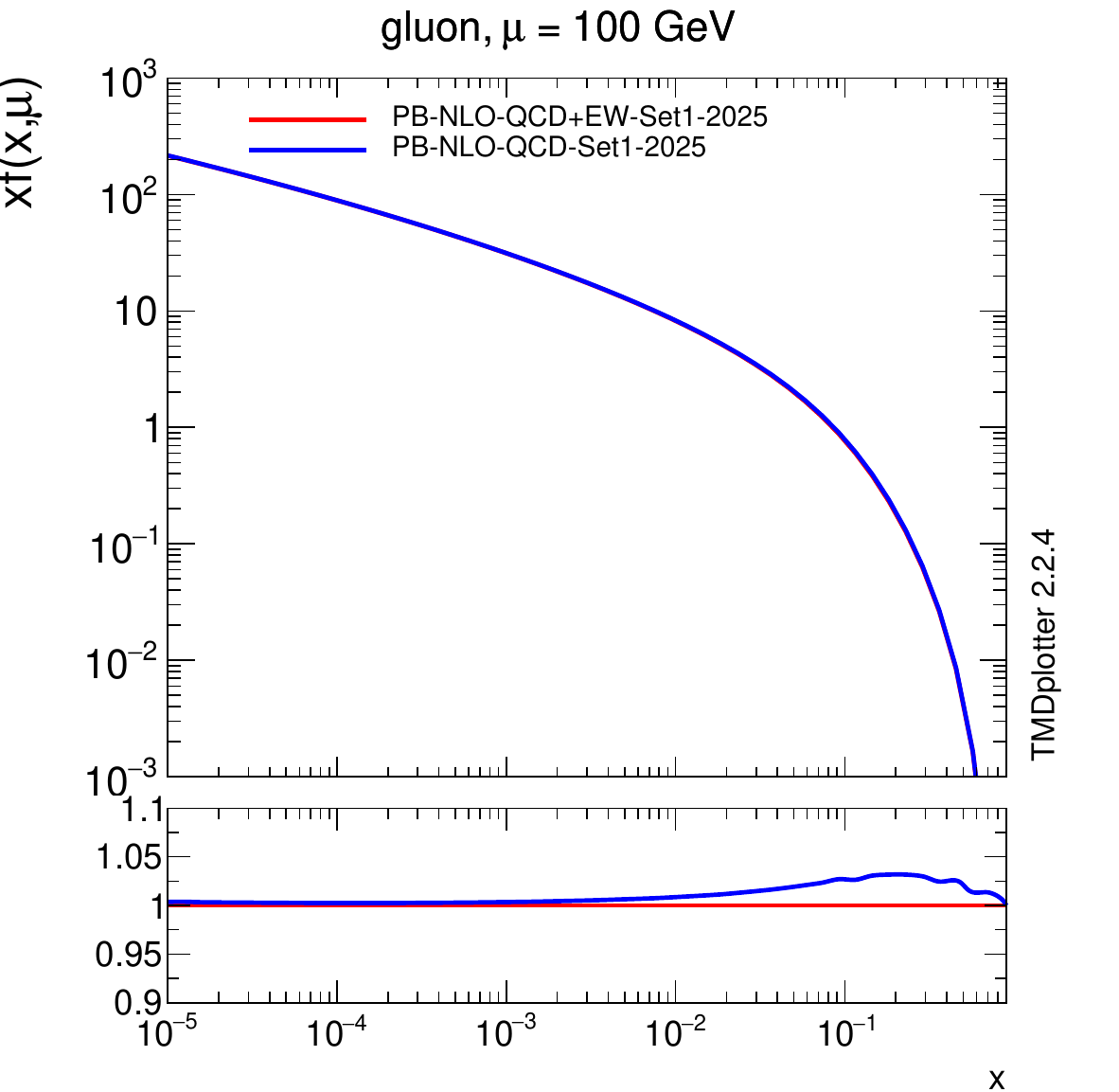}
\caption{\small The collinear quark and gluon densities for Set1 at $\mu= 100$ GeV as a function of $x$. The EW set together with the one containing only QCD partons is shown. The lower panel shows the ratio between both.}
\label{CollPdfFit-Set1}
\end{center}
\end{figure}
\begin{figure}[h!tb]

\begin{center} 
\includegraphics[width=0.32\textwidth,angle=0]{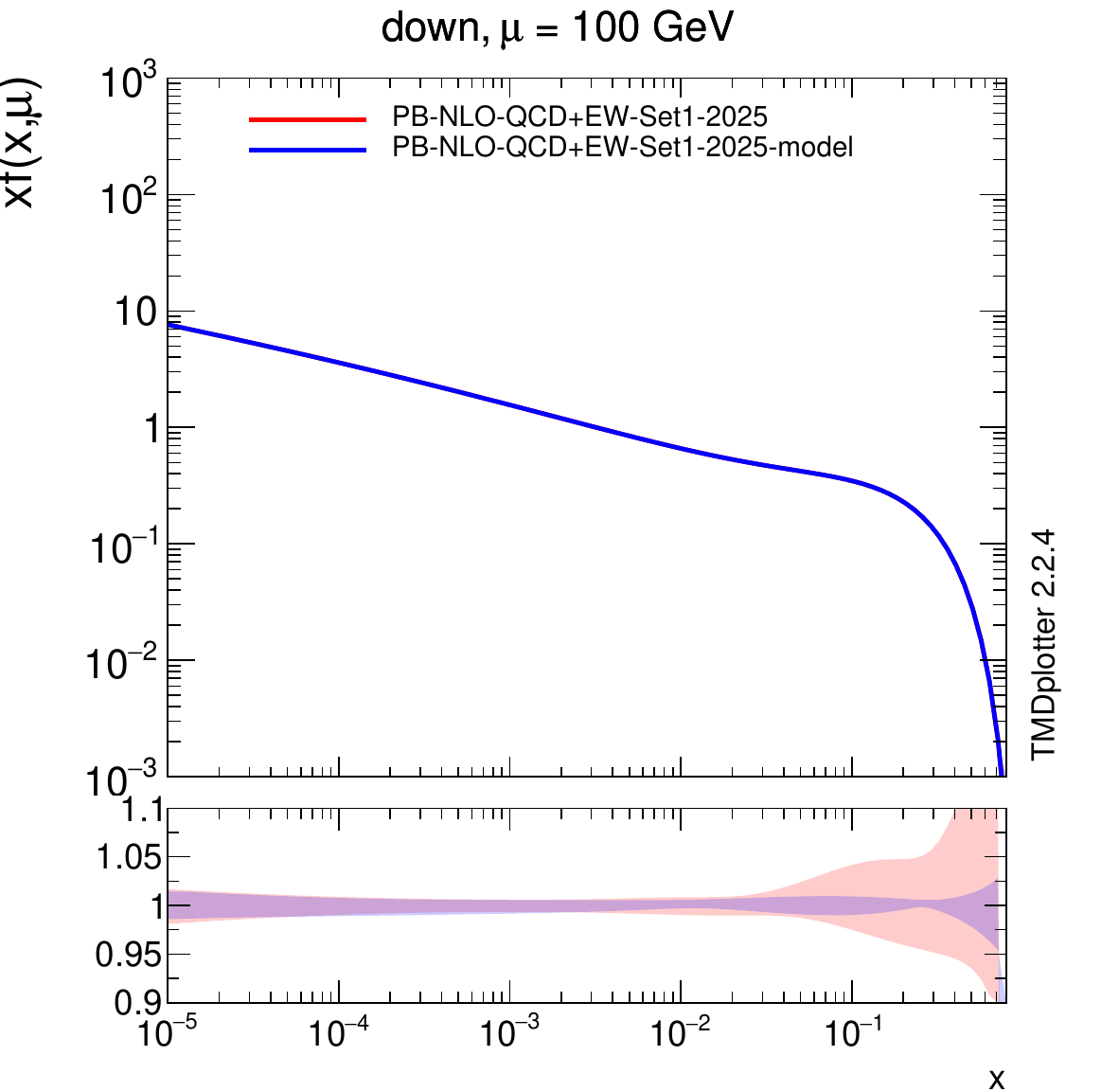}
\includegraphics[width=0.32\textwidth,angle=0]{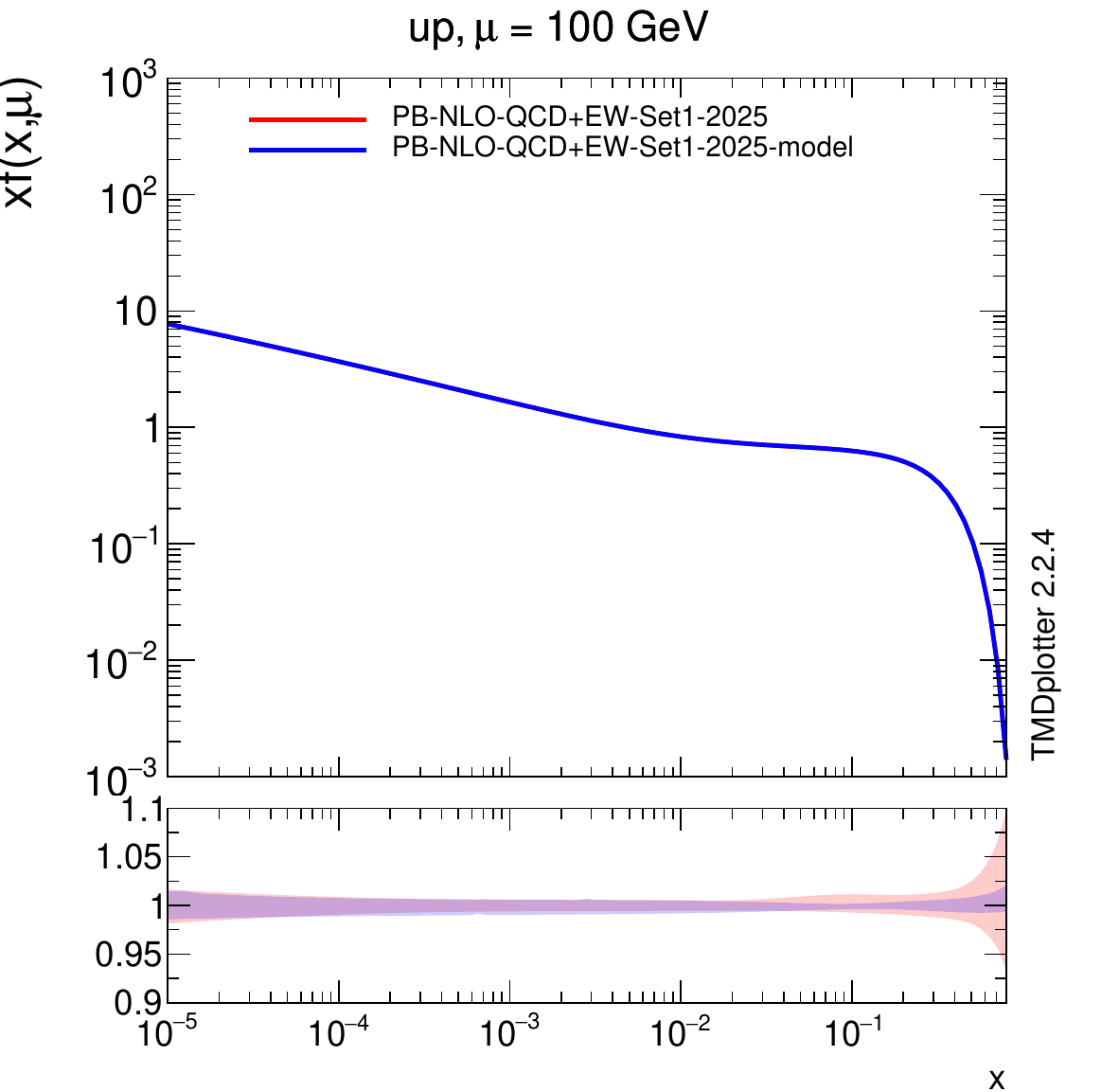}
\includegraphics[width=0.32\textwidth,angle=0]{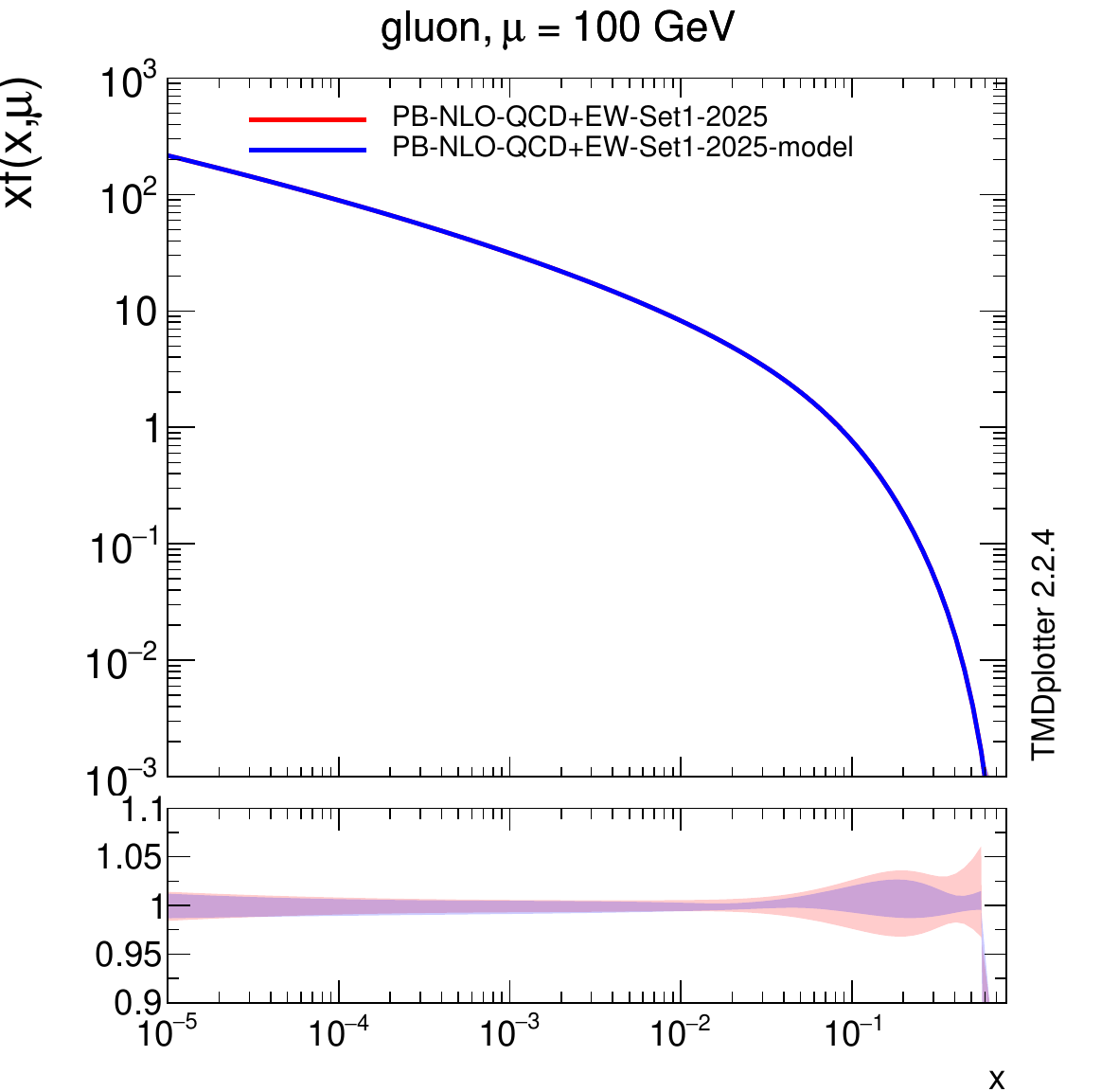}
\includegraphics[width=0.32\textwidth,angle=0]{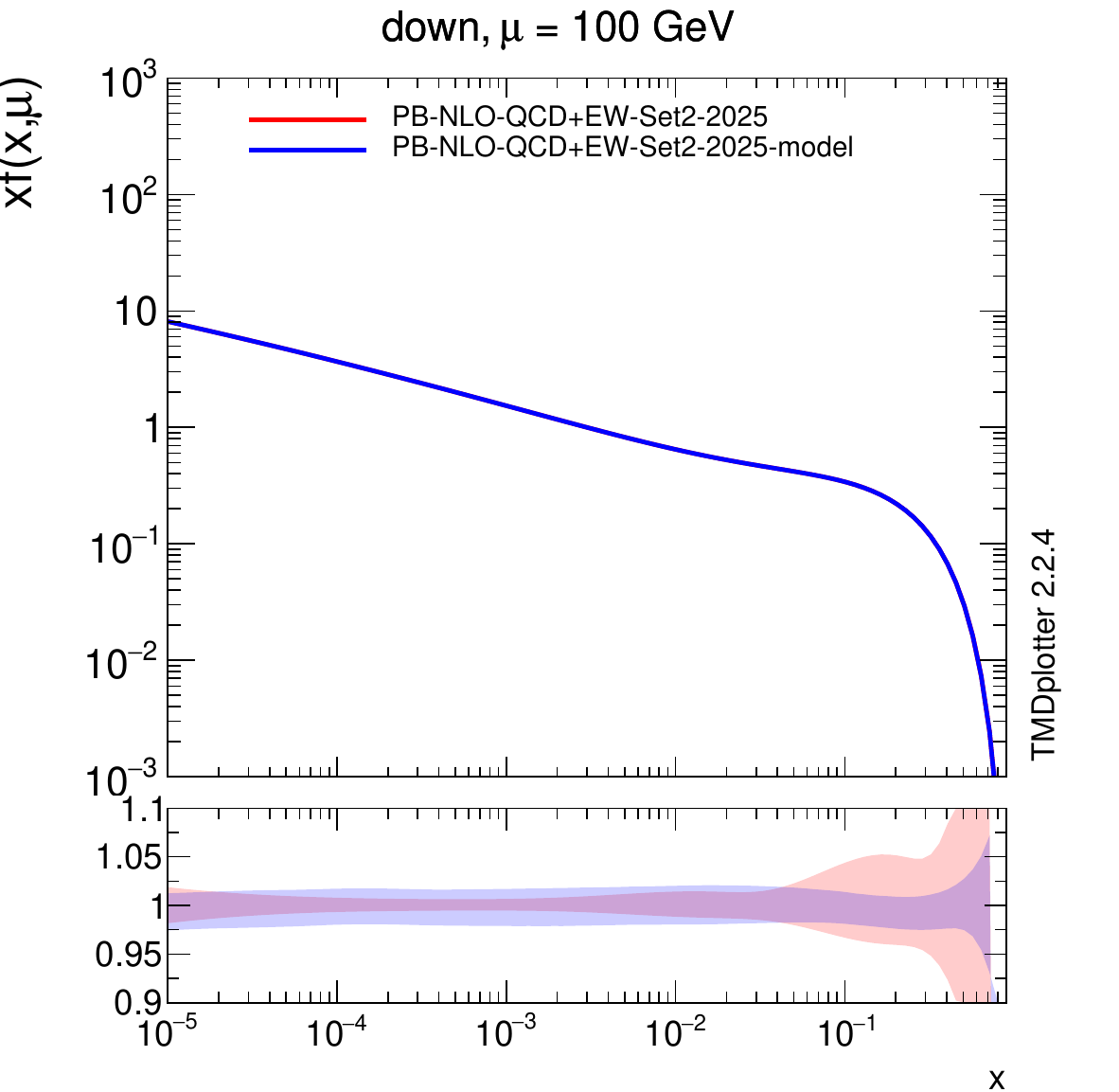}
\includegraphics[width=0.32\textwidth,angle=0]{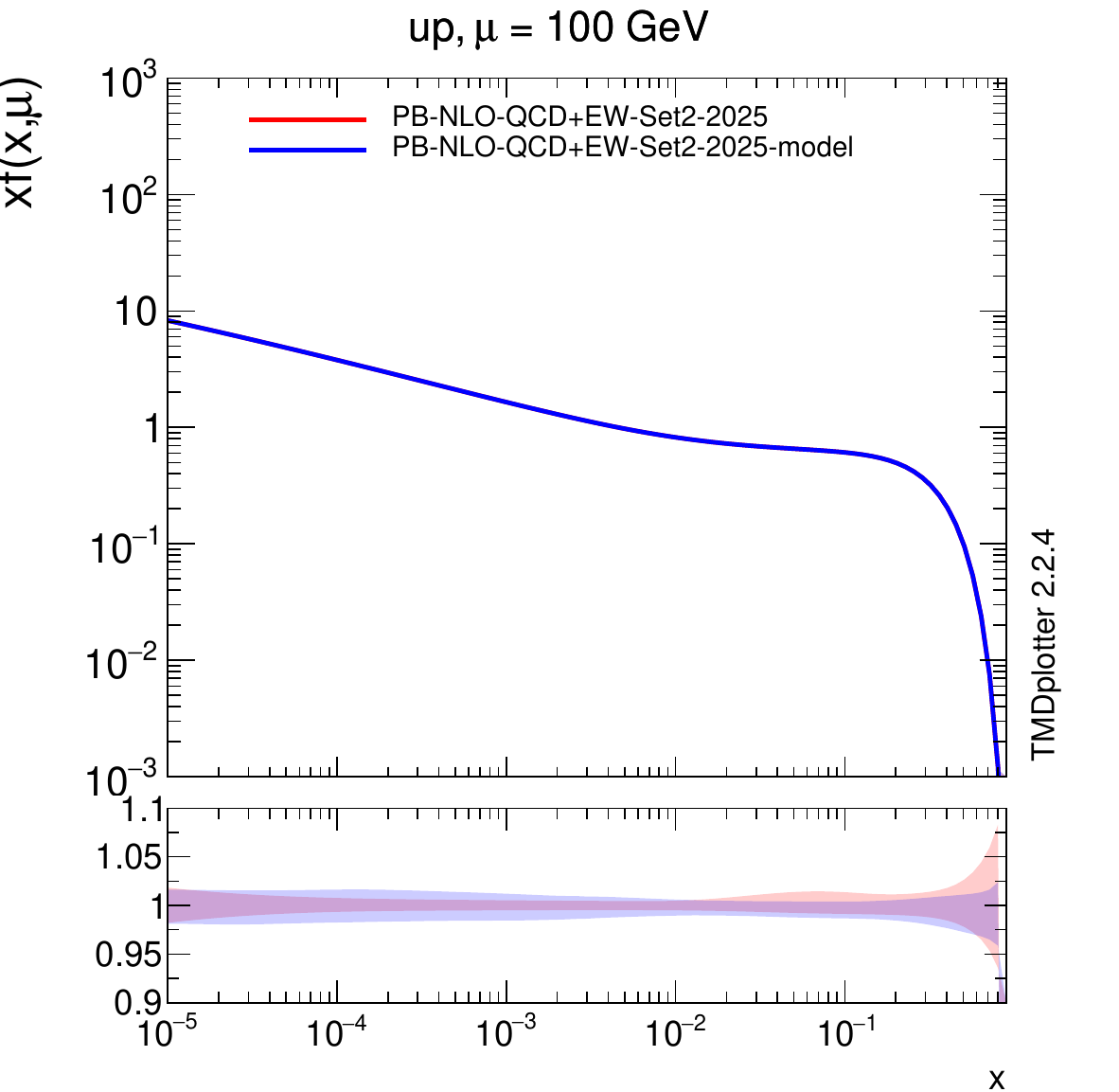}
\includegraphics[width=0.32\textwidth,angle=0]{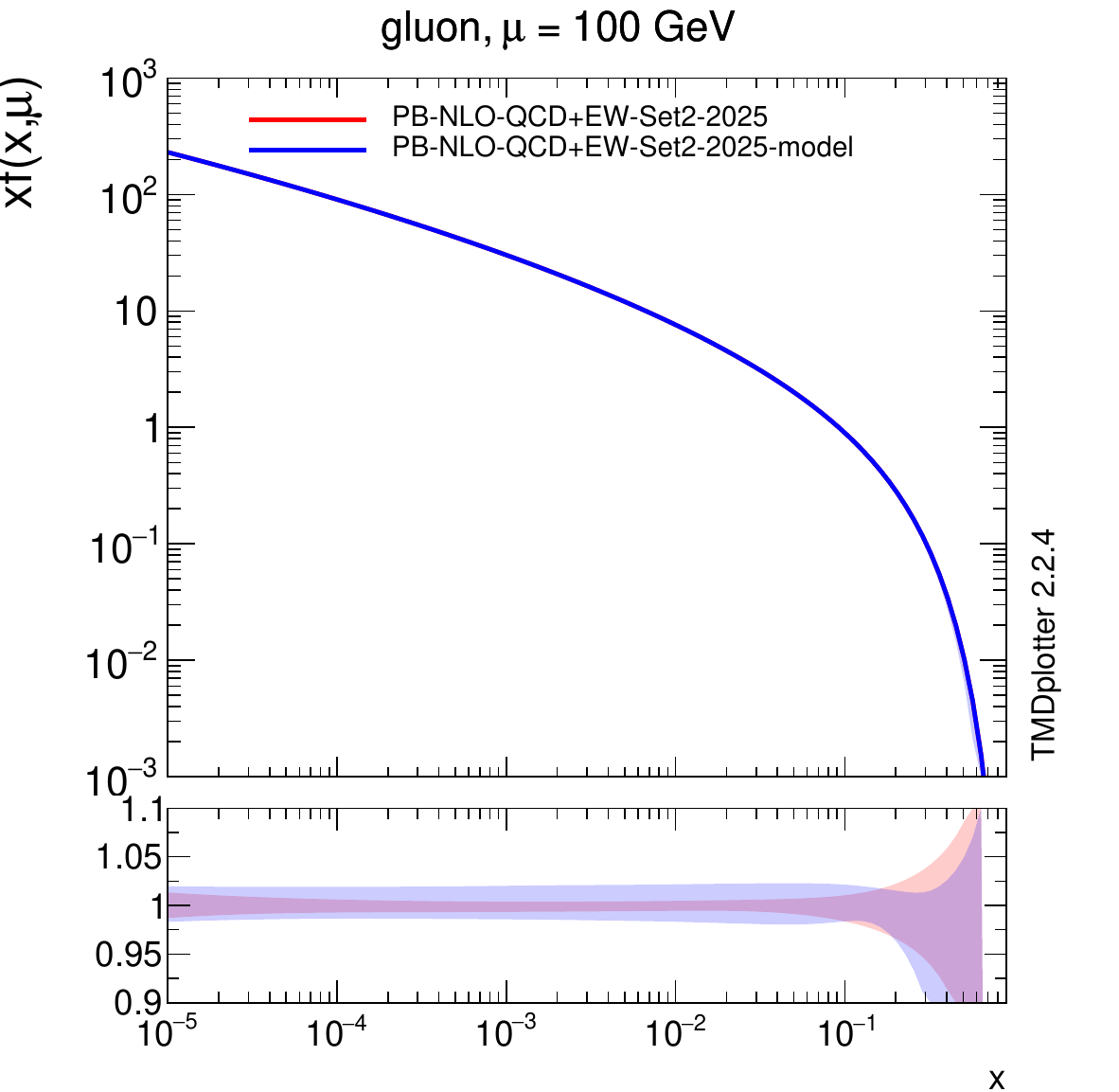}
\caption{\small The collinear quark and gluon densities  at $\mu= 100$ GeV as a function of $x$. Shown is \PBnewEW~Set1 (upper row) and  \PBnewEW~Set2 (lower row). The uncertainty bands for experimental and model uncertainties are shown separately. }
\label{CollPdfFit-Set2}
\end{center}
\end{figure}

In Fig.~\ref{CollPdfFit-Set1} we compare collinear quark and gluon densities at $\mu = 100$~\GeV\ obtained from the full set (including QCD and EW effects, \PBnewEW  ) to the one with QCD partons  only (\PBnewQCD ).
The distributions, including electroweak bosons, are slightly different compared to those containing only QCD partons. The uncertainties shown in Fig.~\ref{CollPdfFit-Set2} do not fully cover these differences.

In Fig.~\ref{CollPdfFit-Set2} we compare distributions obtained from \PBnewEW\  Set1 and Set2. Uncertainties, obtained as described in Sec.~\ref{uncertainties}, are shown in the lower panels. At small $x$ there are small differences in the quark distributions, the gluon distribution differs at large $x$. These differences come from the different treatment of the scale used in \as : in Set 1 the evolution scale $q$  is used, while in Set2 the transverse momentum \qt\ is applied.
In Fig.~\ref{TMDFit-Set1+2} in Appendix~\ref{TMDs} we show the TMD distributions for quarks and gluons at $\mu= 100$ GeV and $x=0.01$.

\subsection{Photon densities}

\begin{figure}[h!tb]
\begin{center} 
\includegraphics[width=0.32\textwidth,angle=0]{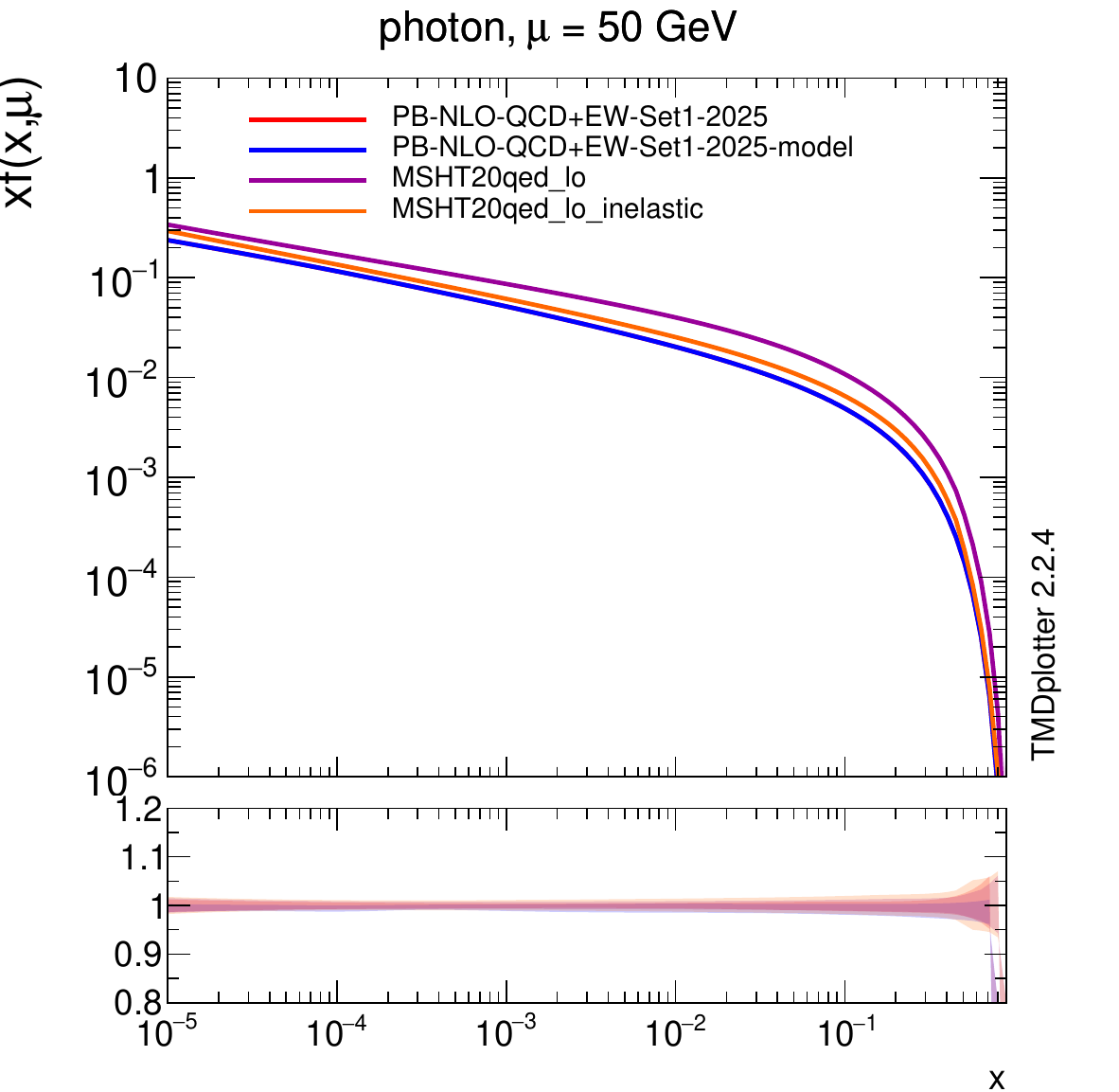}
\includegraphics[width=0.32\textwidth,angle=0]{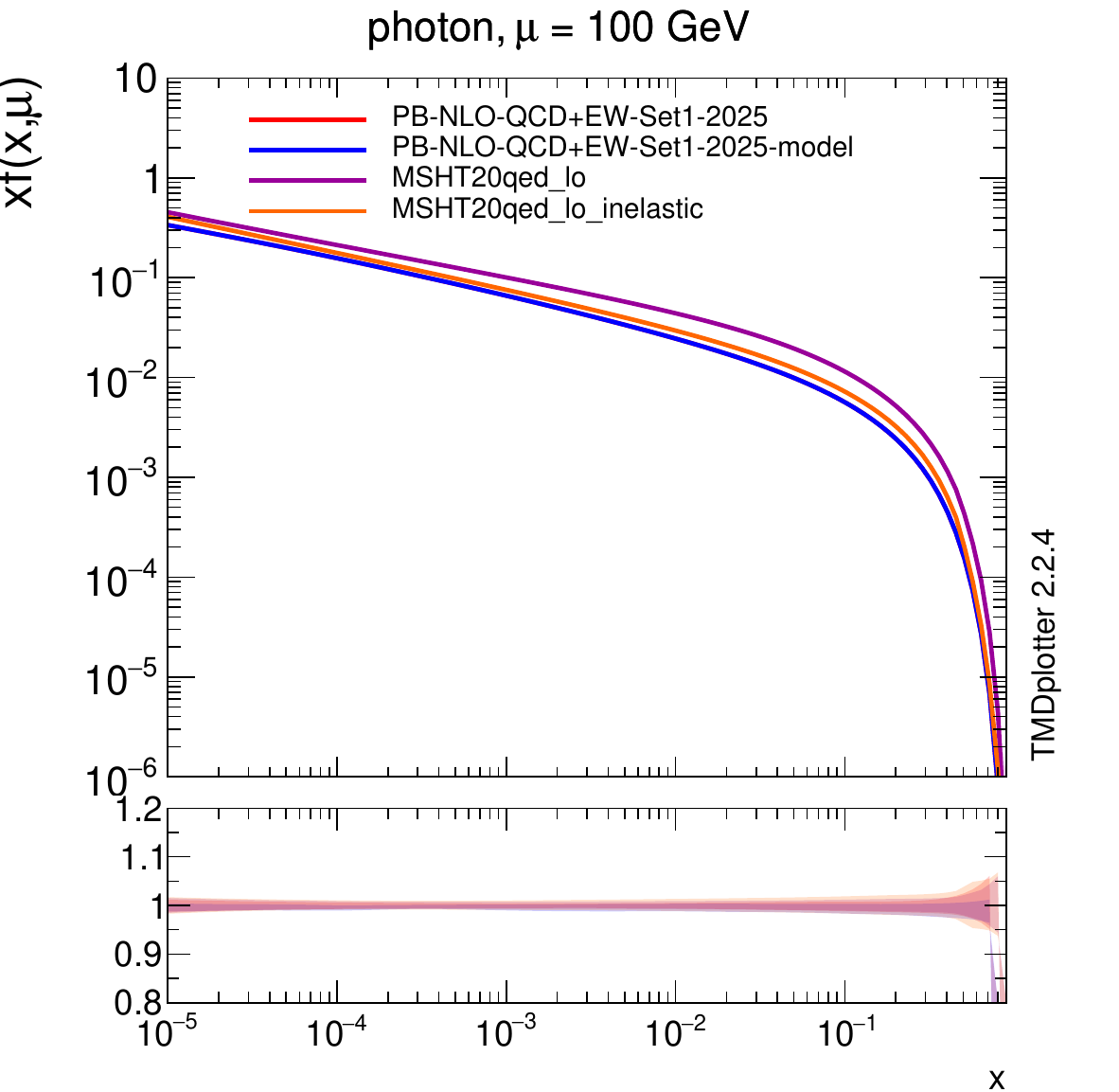}
\includegraphics[width=0.32\textwidth,angle=0]{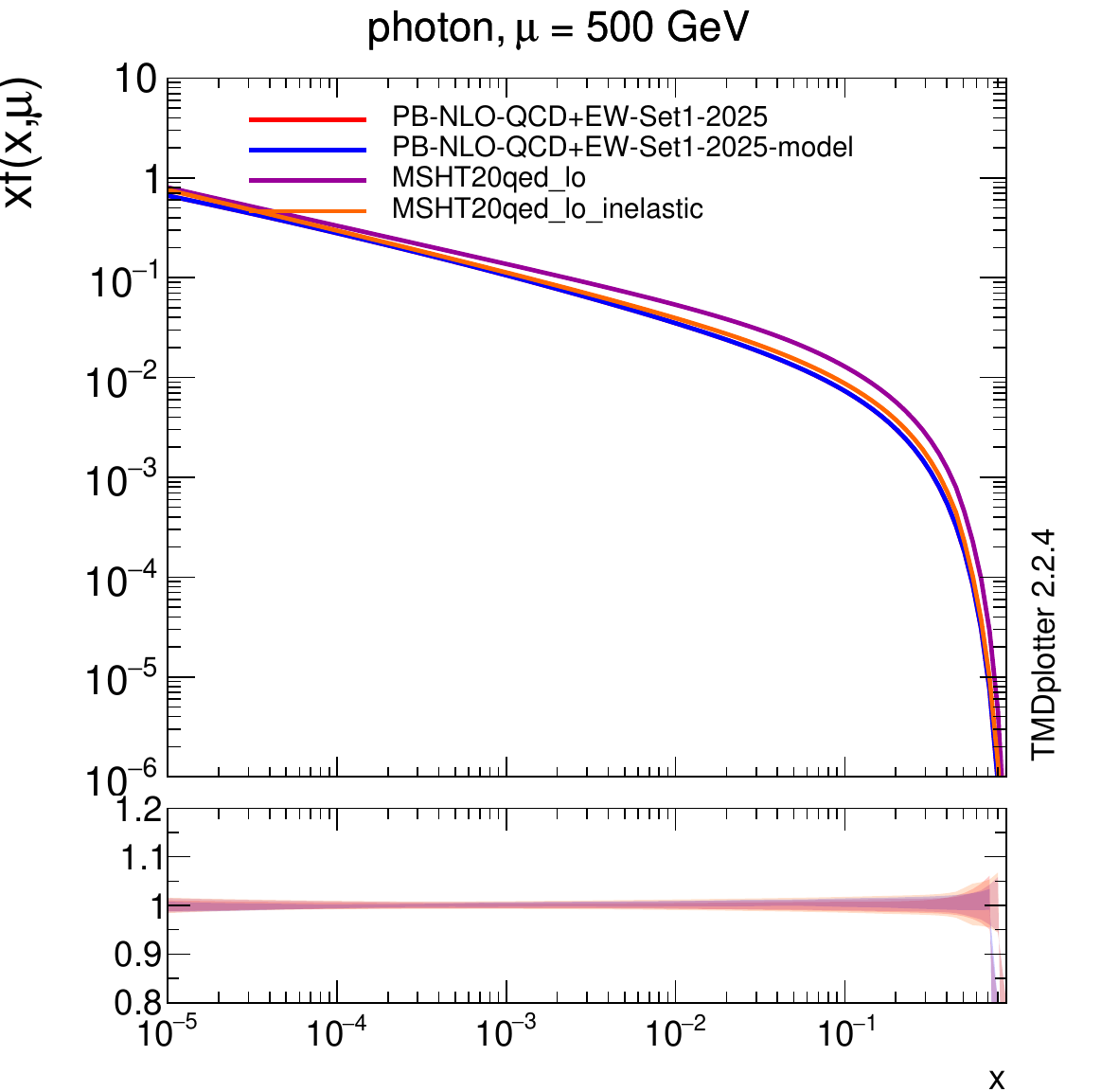}
\includegraphics[width=0.32\textwidth,angle=0]{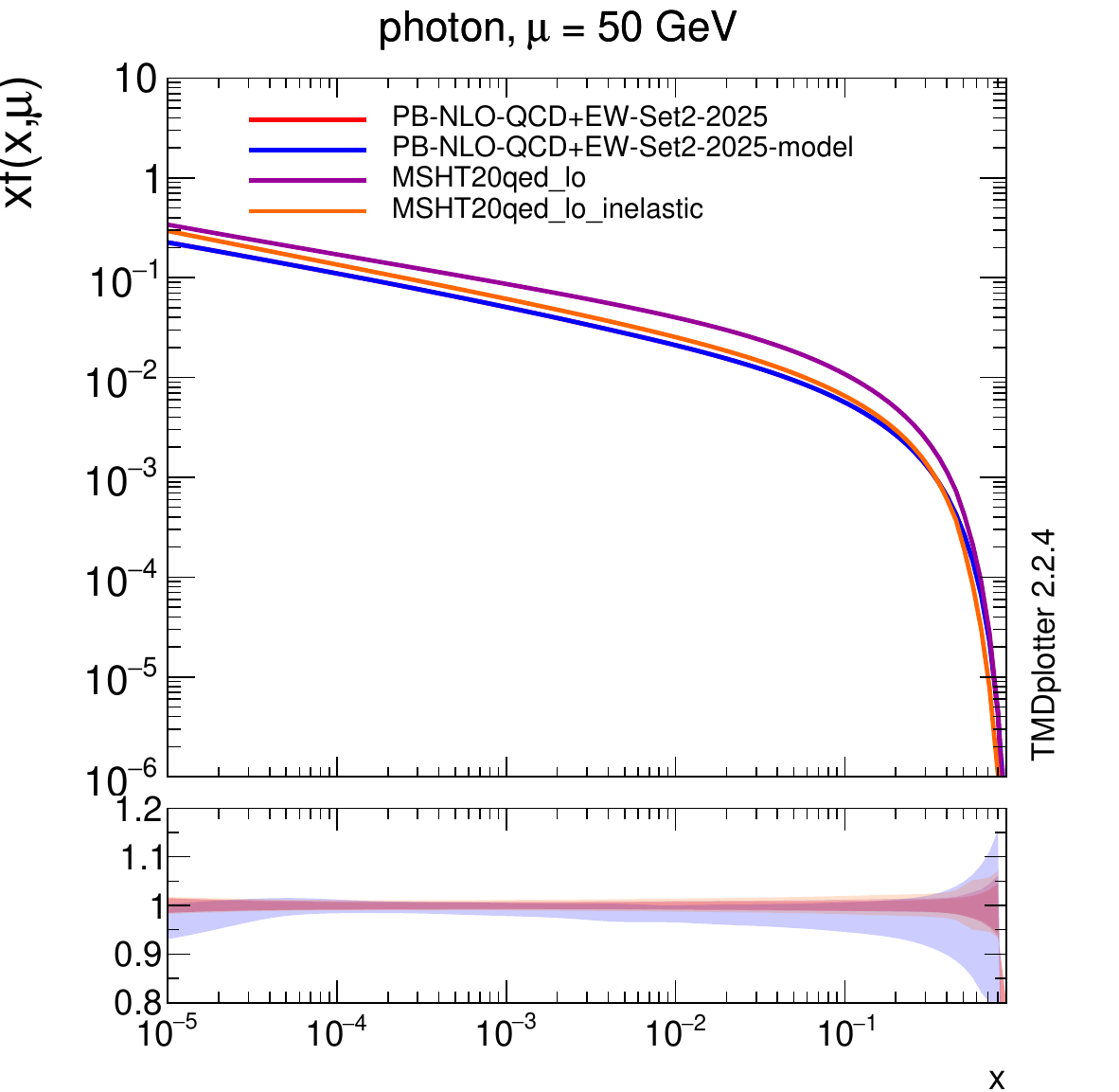}
\includegraphics[width=0.32\textwidth,angle=0]{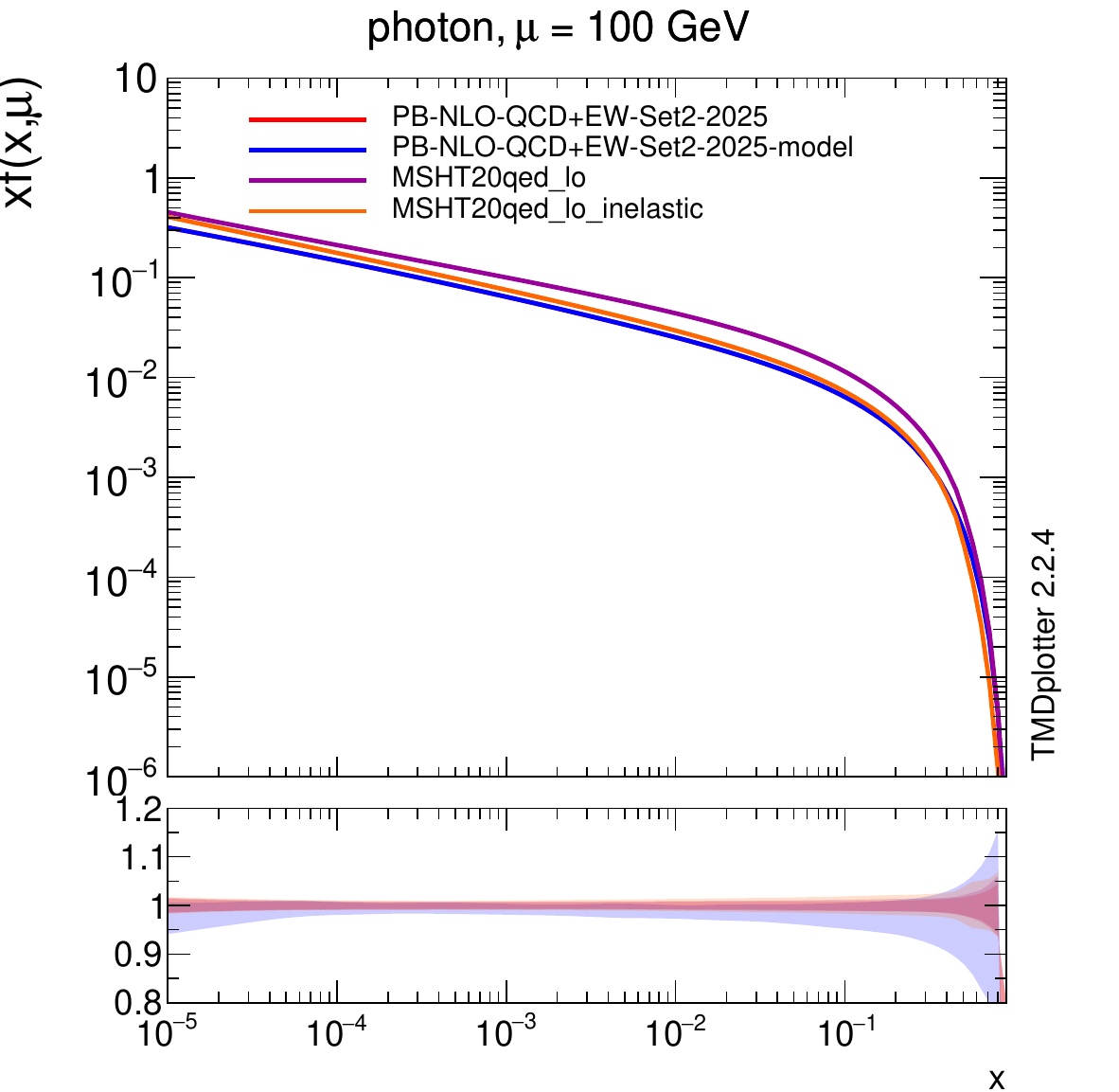}
\includegraphics[width=0.32\textwidth,angle=0]{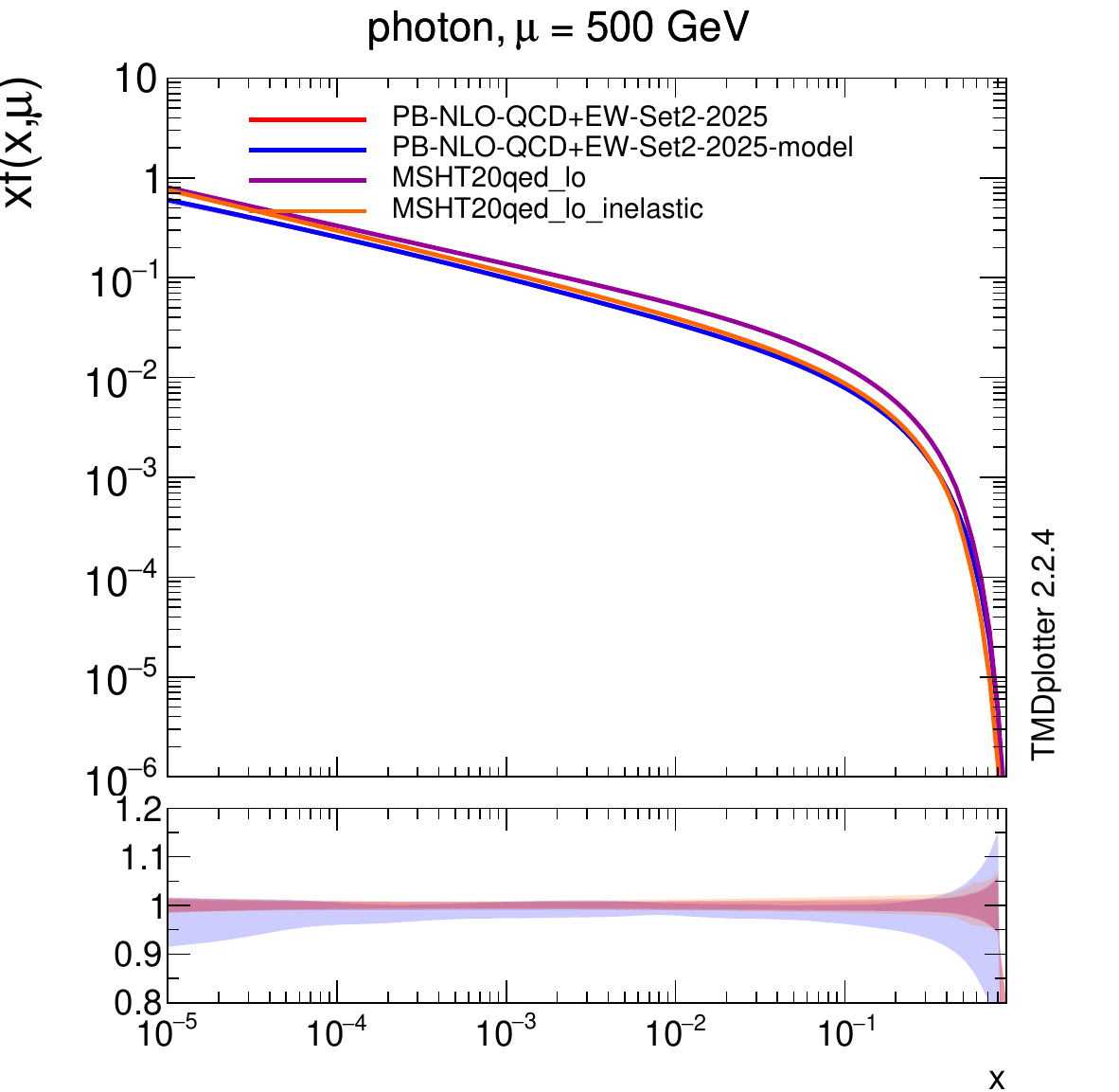}
\caption{\small The collinear photon density for  \PBnewEW~Set1 (upper row)  and Set2 (lower row)  at $\mu= 50$, 100 and 500~GeV  as a function of $x$ including experimental and model uncertainties. 
For comparison  the prediction from MSHT20~\protect\cite{Cridge:2021pxm} (total and inelastic) is shown.  
}

\label{PhotonFig1}
\end{center}
\end{figure}

The collinear photon density, obtained from the fit\footnote{There are small differences to the one published in Ref.~\cite{Jung:2021mox} due to mistreatment of the number of flavors.}, as a function of $x$  for different values of the evolution scales $\mu$ is shown in Fig.~\ref{PhotonFig1} for  \PBnewEW~Set1 and  \PBnewEW~Set2. 
The uncertainties shown come from the experimental and model uncertainties, as described in Sec.~\ref{uncertainties}. 
For comparison the MSHT20qed photon densities~\cite{Cridge:2021pxm} are shown. The inelastic component, comparable with the photon density described in this paper is shown separately. Rather good agreement is observed, the small differences come from $\alpha_{em}$, which is taken as $\alpha_{em}=\alpha_0=1/137$ for consistency with the electroweak parameters.

In Fig.~\ref{PhotonFig2} of Appendix~ \ref{TMDs} the TMD density of photons is shown.

\subsection{Heavy Boson densities\label{EWdensities}}
\begin{figure}[h!tb]
\begin{center} 
\includegraphics[width=0.32\textwidth,angle=0]{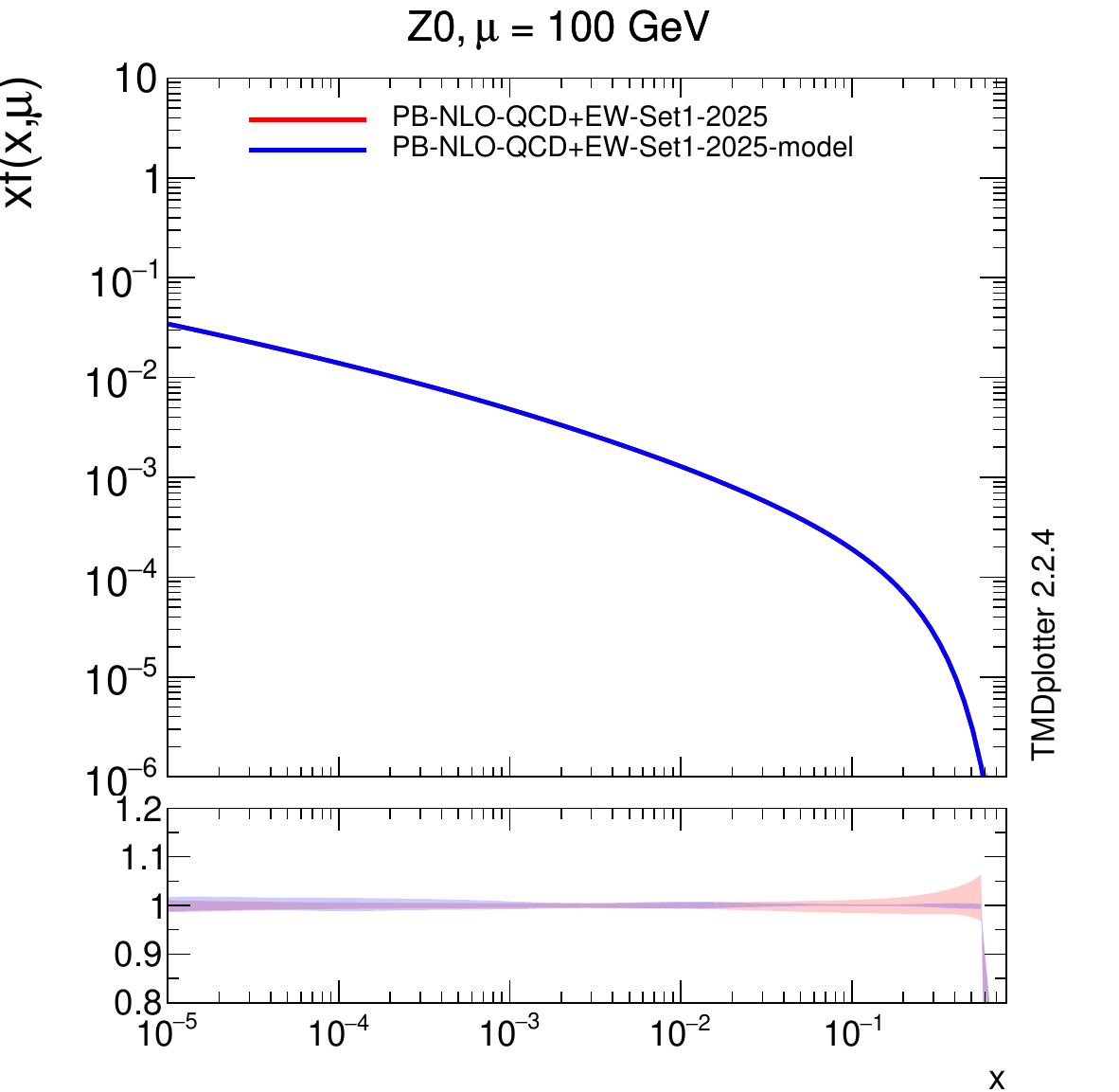}
\includegraphics[width=0.32\textwidth,angle=0]{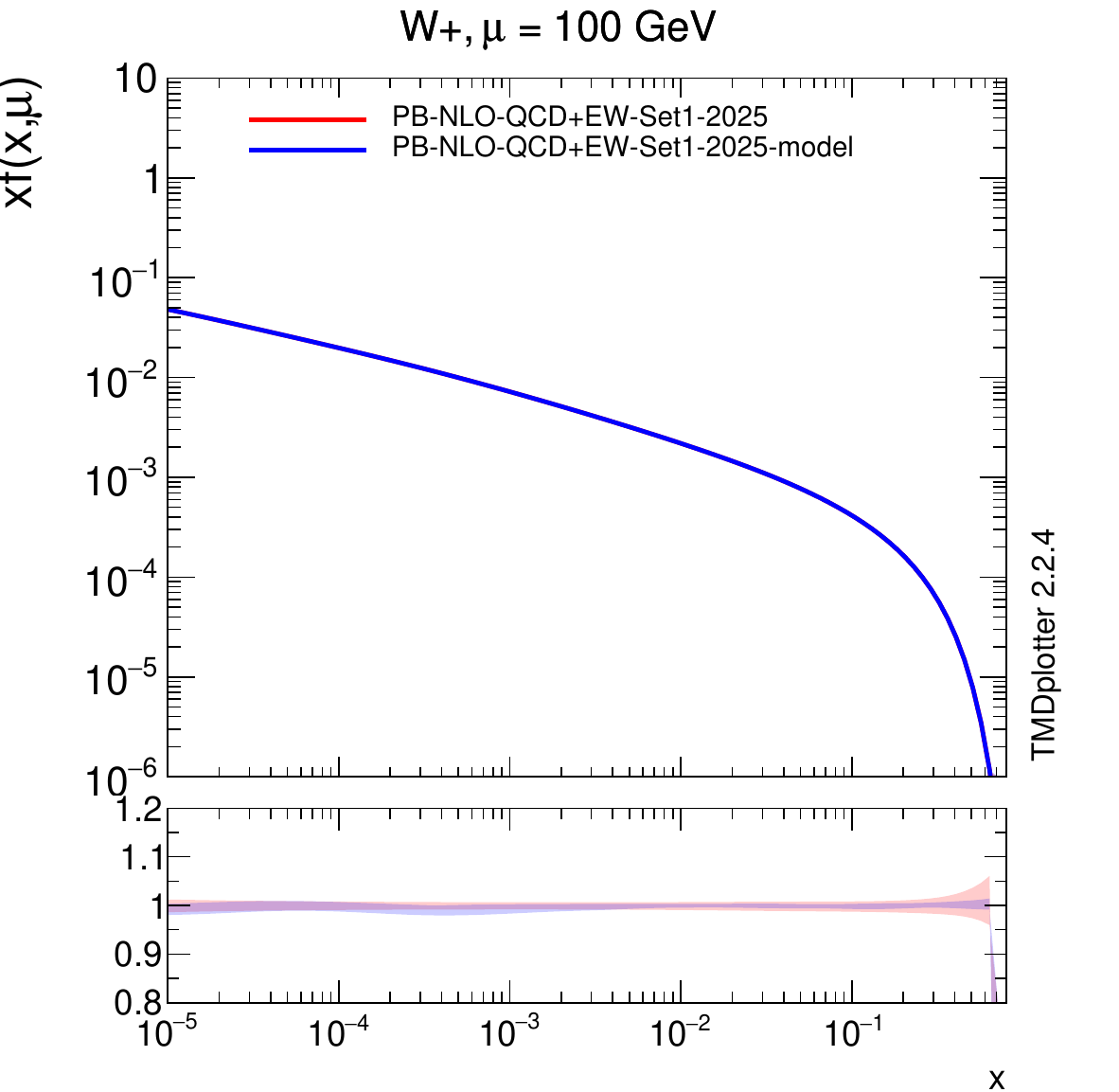}
\includegraphics[width=0.32\textwidth,angle=0]{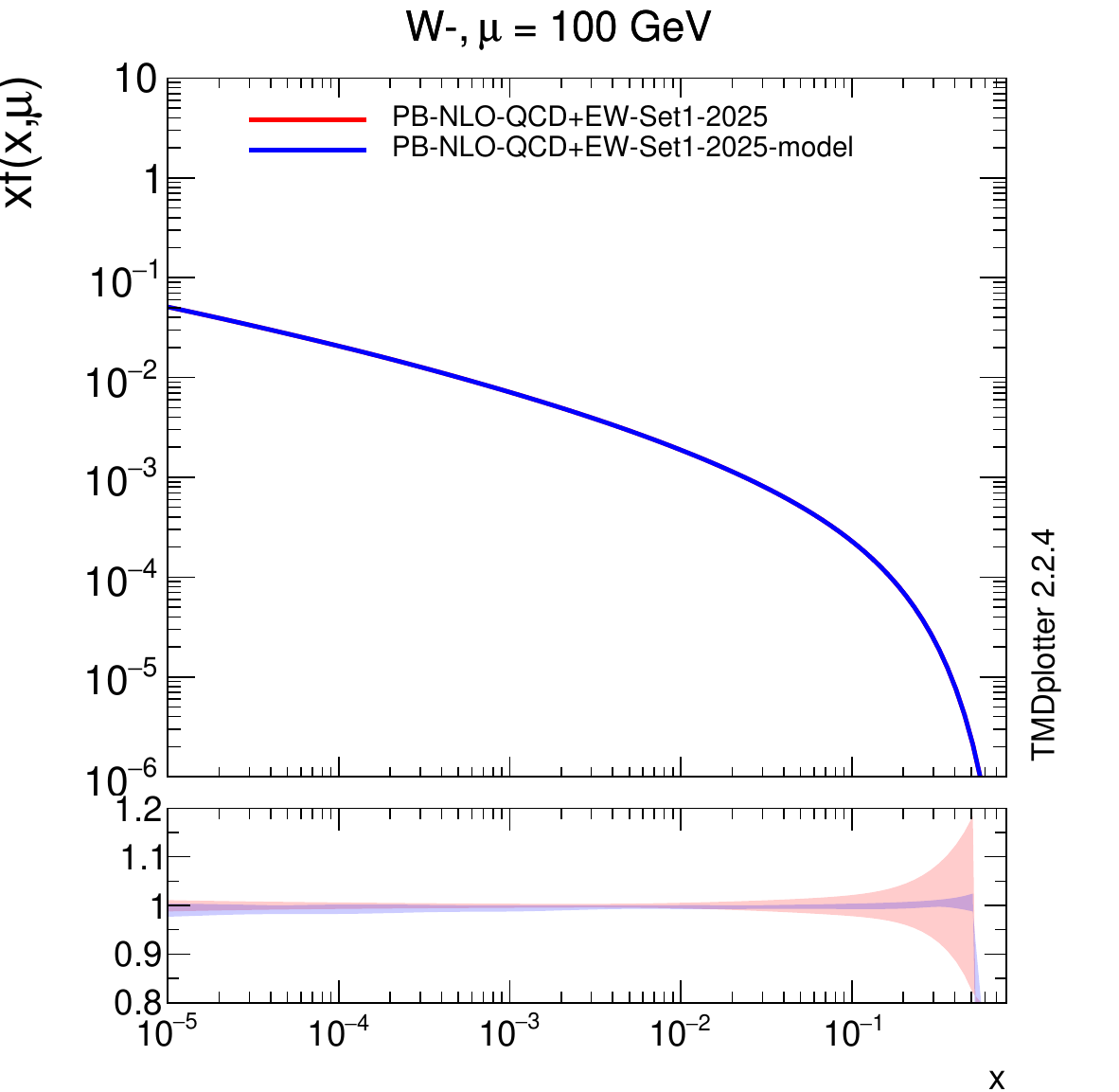}
\caption{\small The collinear vector-boson densities  for  \PBnewEW~Set1 at $\mu= 100$~GeV  as a function of $x$ including experimental and model uncertainties. }
\label{CollEWSet1}
\end{center}
\end{figure}
\begin{figure}[h!tb]
\begin{center} 
\includegraphics[width=0.32\textwidth,angle=0]{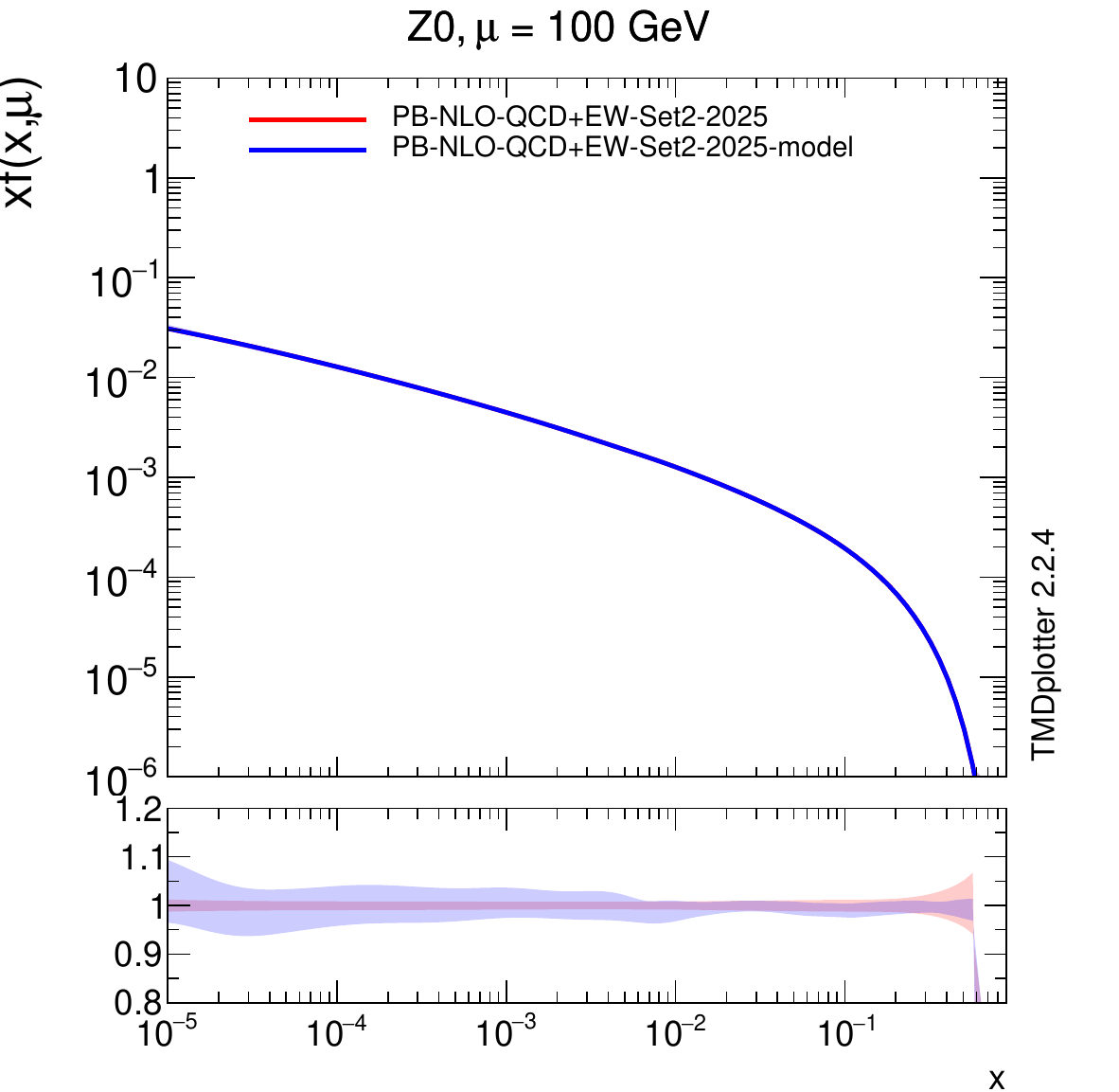}
\includegraphics[width=0.32\textwidth,angle=0]{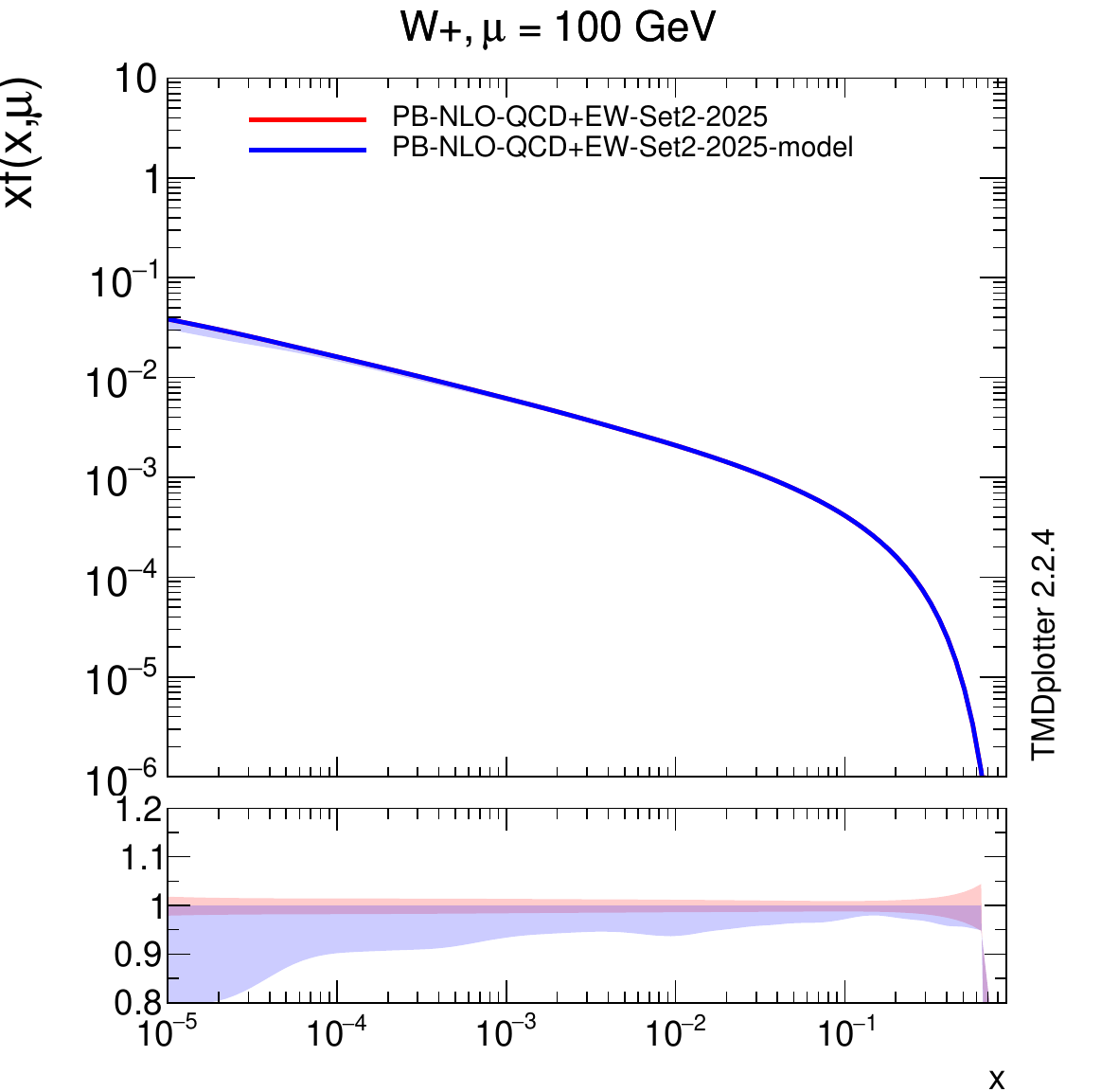}
\includegraphics[width=0.32\textwidth,angle=0]{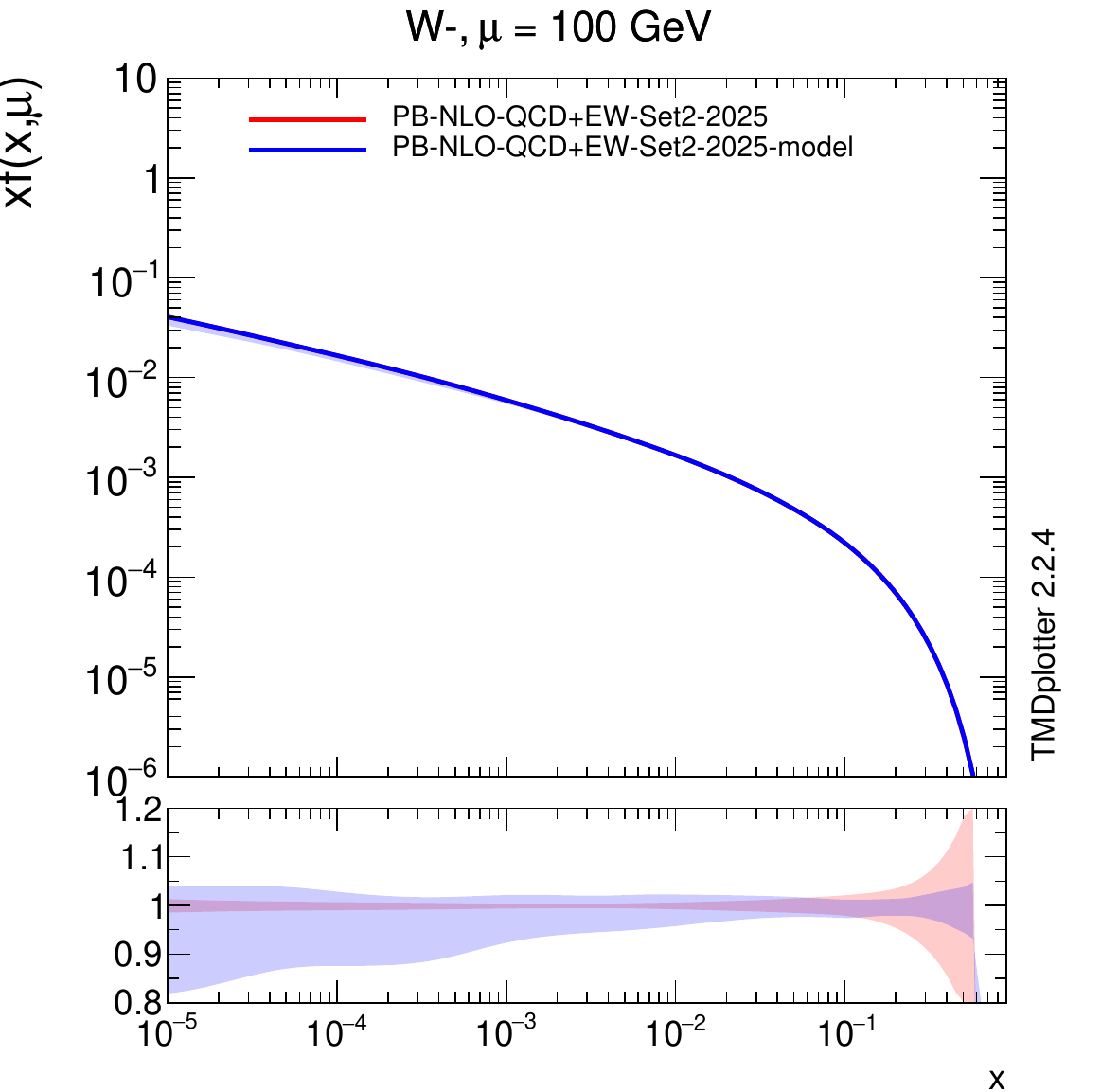}
\caption{\small The collinear vector-boson density for  \PBnewEW~Set2 at $\mu= 100$~GeV  as a function of $x$ including experimental and model uncertainties. }
\label{CollEWSet2}
\end{center}
\end{figure}

The approach to determine the photon densities within the \PB -method has been extended to determine the collinear and TMD densities of \PZ\ and \PW\  as described in eqs.(\ref{ZPDF}-\ref{W+PDF}). The  collinear  densities for \PZ\ and \PW\ are shown in Figs.~\ref{CollEWSet1} and~\ref{CollEWSet2} using {\sc MassCutScheme=1}. 
The $\PW^+$ and $\PW^-$ densities are different, because of the different couplings to u- and d-quarks. The uncertainties shown come from the experimental uncertainties as well as model uncertainties as described in Sec.~\ref{uncertainties}. The uncertainties coming from changing the heavy boson masses are much smaller than all the other uncertainties, and thus are neglected. The model uncertainties  result from the large $x$-model dependence of the quark distributions, which can become significant. 

The momentum fractions carried by photons, \PZ\ and \PW - bosons from $\int_0^1 dx xf(x,\mu) $ at $\mu=100$ \GeV$^2$ are shown in Tab.~\ref{MomFracl}.
\begin{table}[htb]
\renewcommand*{\arraystretch}{1.0}
\centerline{
\begin{tabular}{|c|c|c|c|}
\hline
{\small $\int_0^1 dx \, xf_i(x,\mu) $ }&
{\footnotesize \PBnewEW~Set1} &
{\footnotesize \PBnewEW~Set2}  & 
{\footnotesize MSHT20qed\_lo\_inelastic} \\
\hline
$i=\Pgamma$ & 0.0021 &  0.0023 &  0.0026\\
$i=\PZ $          & $8.7 \cdot 10^{-5}$ & $8.6  \cdot 10^{-5}$ & - \\
$i=\PW ^+$  & $ 1.7 \cdot 10^{-4}$& $ 1.6 \cdot 10^{-4}$& - \\
$i=\PW^-$  & $ 1.2 \cdot 10^{-4}$& $ 1.0 \cdot 10^{-4}$  & - \\ \hline
\end{tabular}}
\caption{\small Momentum fraction $\int_0^1 dx \, xf(x,\mu) $ carried by \Pgamma , \PZ , and \PW\ for  \PBnewEW~Set1,    \PBnewEW~Set2  and MSHT20qed\_lo\_inelastic at $\mu= 100$ \GeV} 
\label{MomFracl}
\end{table}

\begin{figure}[h!tb]
\begin{center} 
\includegraphics[width=0.32\textwidth,angle=0]{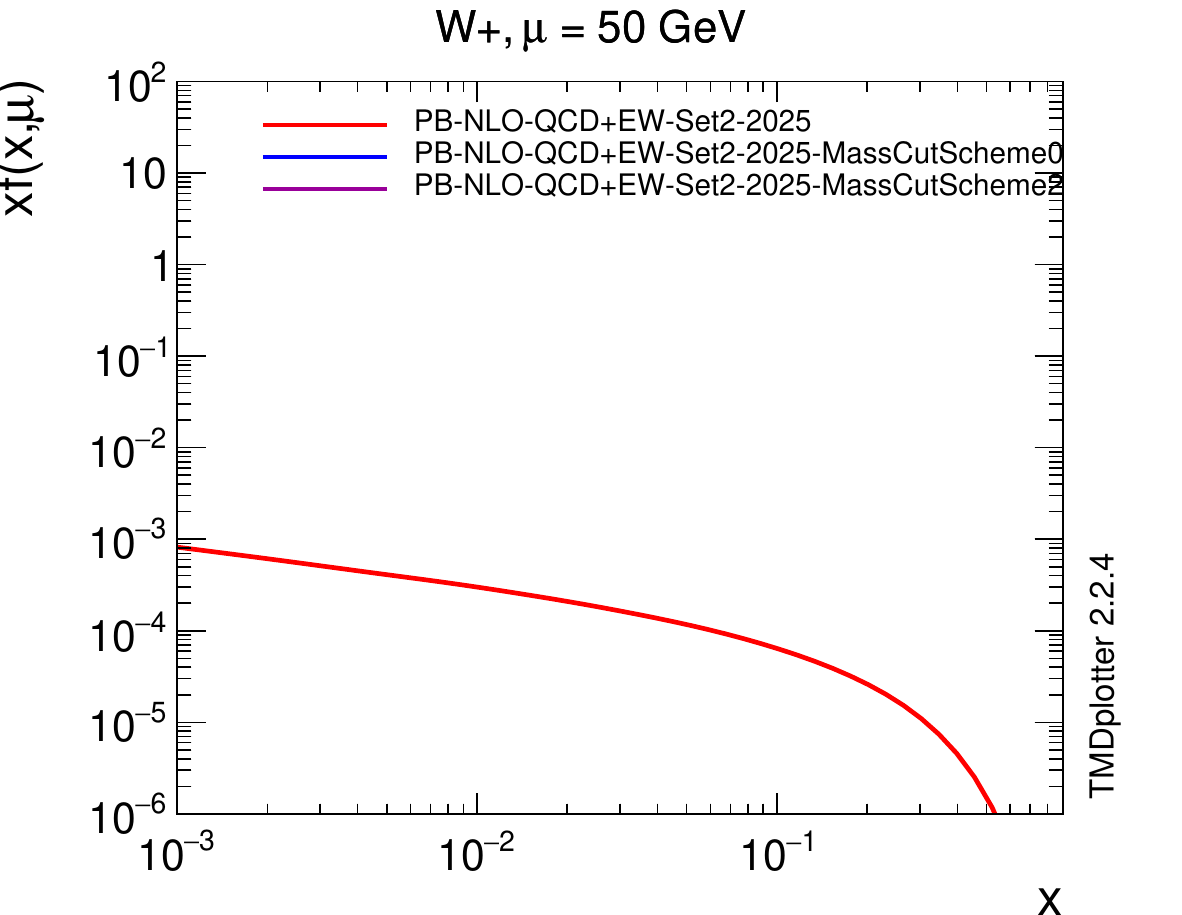}
\includegraphics[width=0.32\textwidth,angle=0]{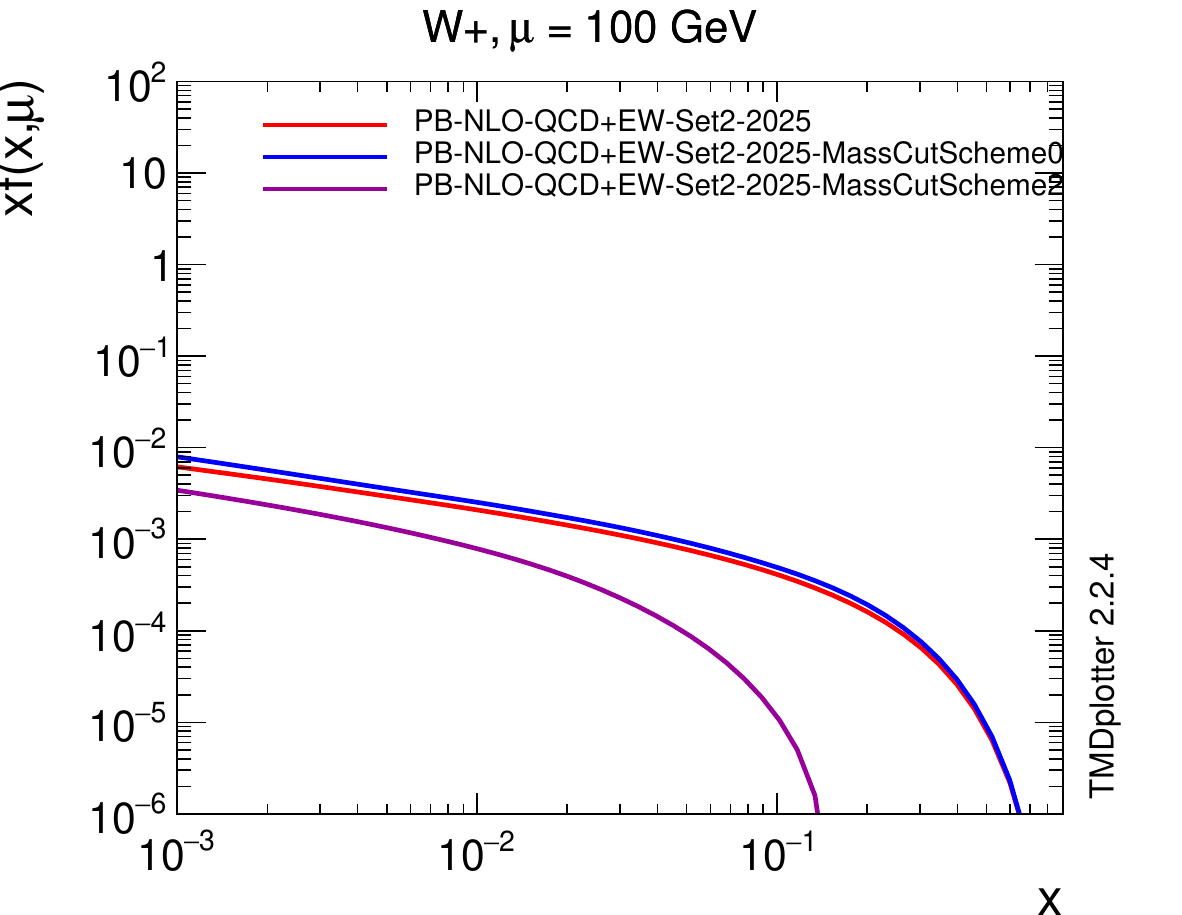}
\includegraphics[width=0.32\textwidth,angle=0]{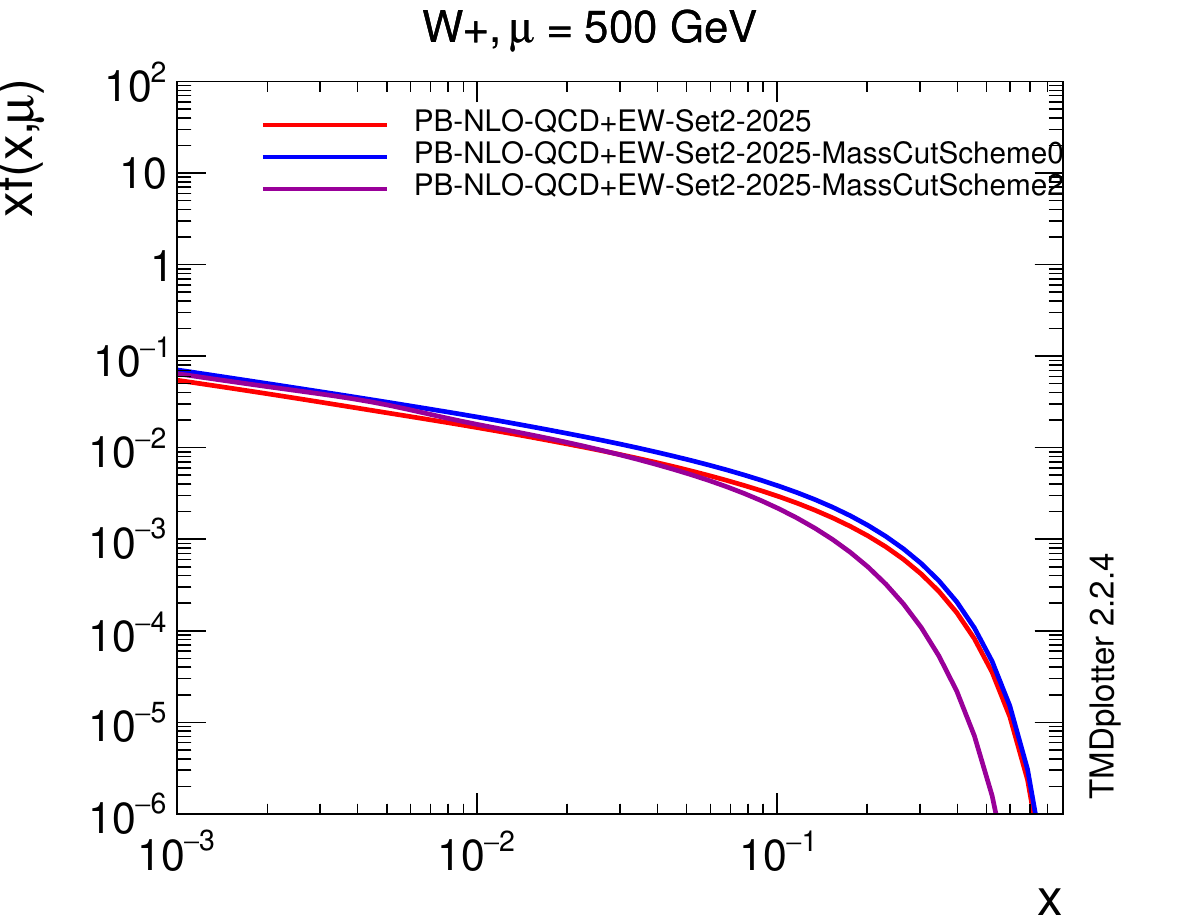}
\caption{\small The collinear density for the \PW -boson   for  \PBnewEW~Set2 at $\mu= 50$, 100  and $\mu= 500$ GeV as a function of $x$ for the different {\sc MassCutSchemes}.}
\label{CollEWFig3}
\end{center}
\end{figure}

\begin{figure}[h!tb]
\begin{center} 
\includegraphics[width=0.32\textwidth,angle=0]{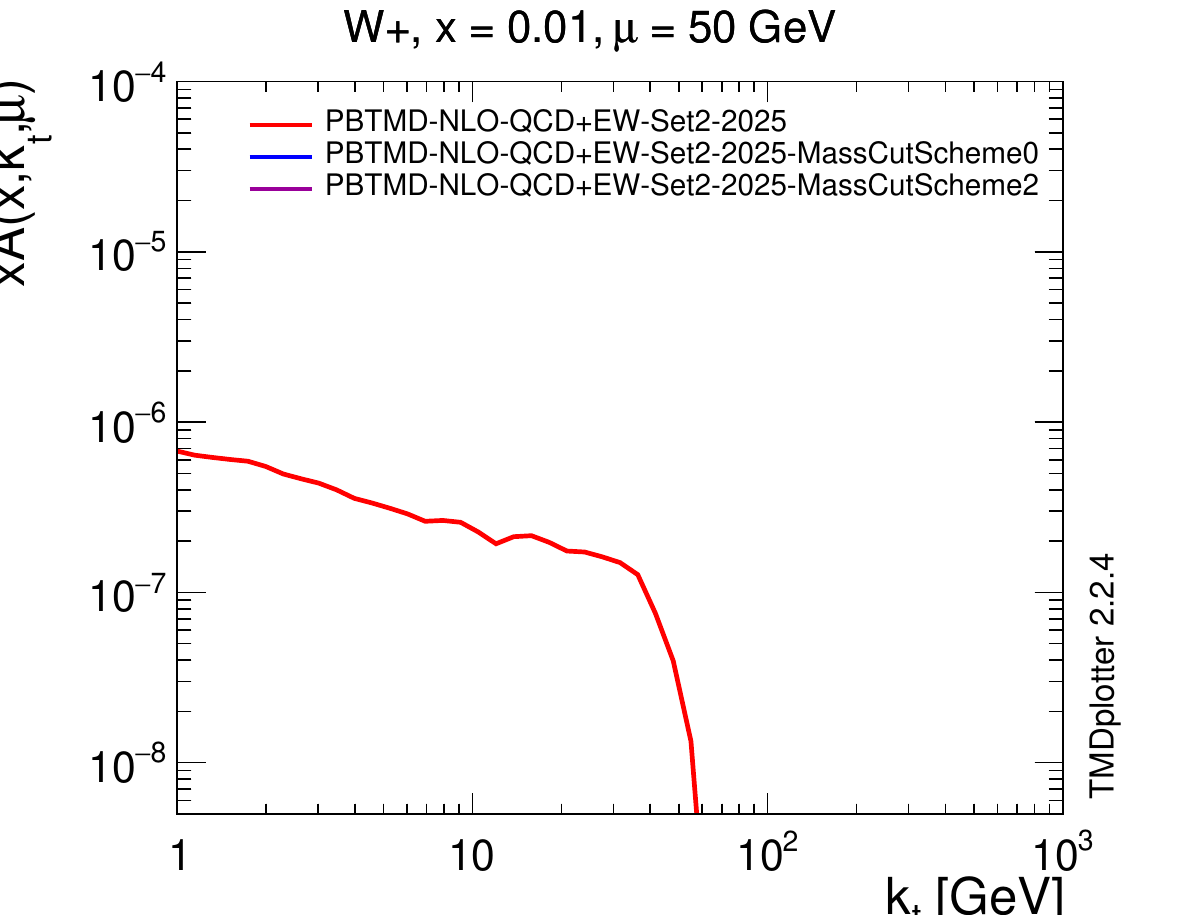}
\includegraphics[width=0.32\textwidth,angle=0]{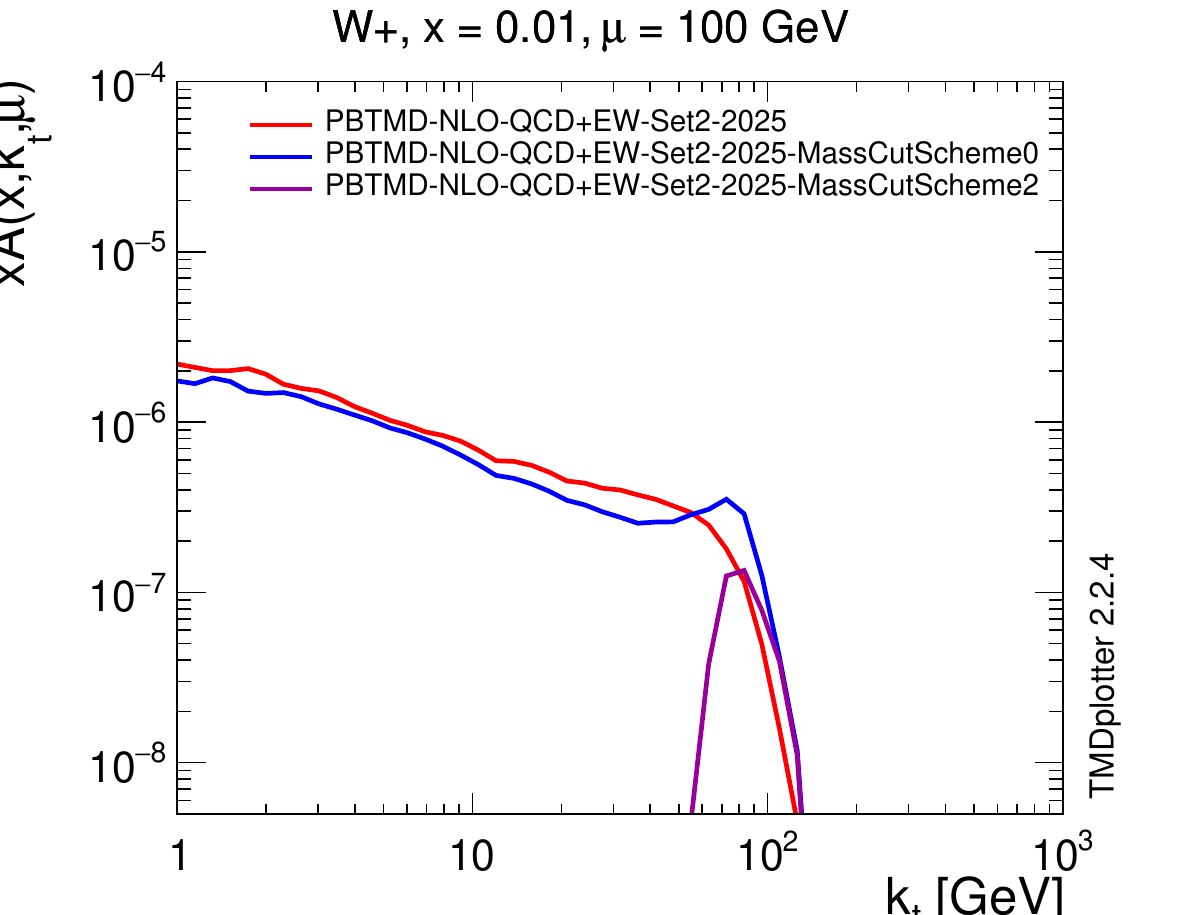}
\includegraphics[width=0.32\textwidth,angle=0]{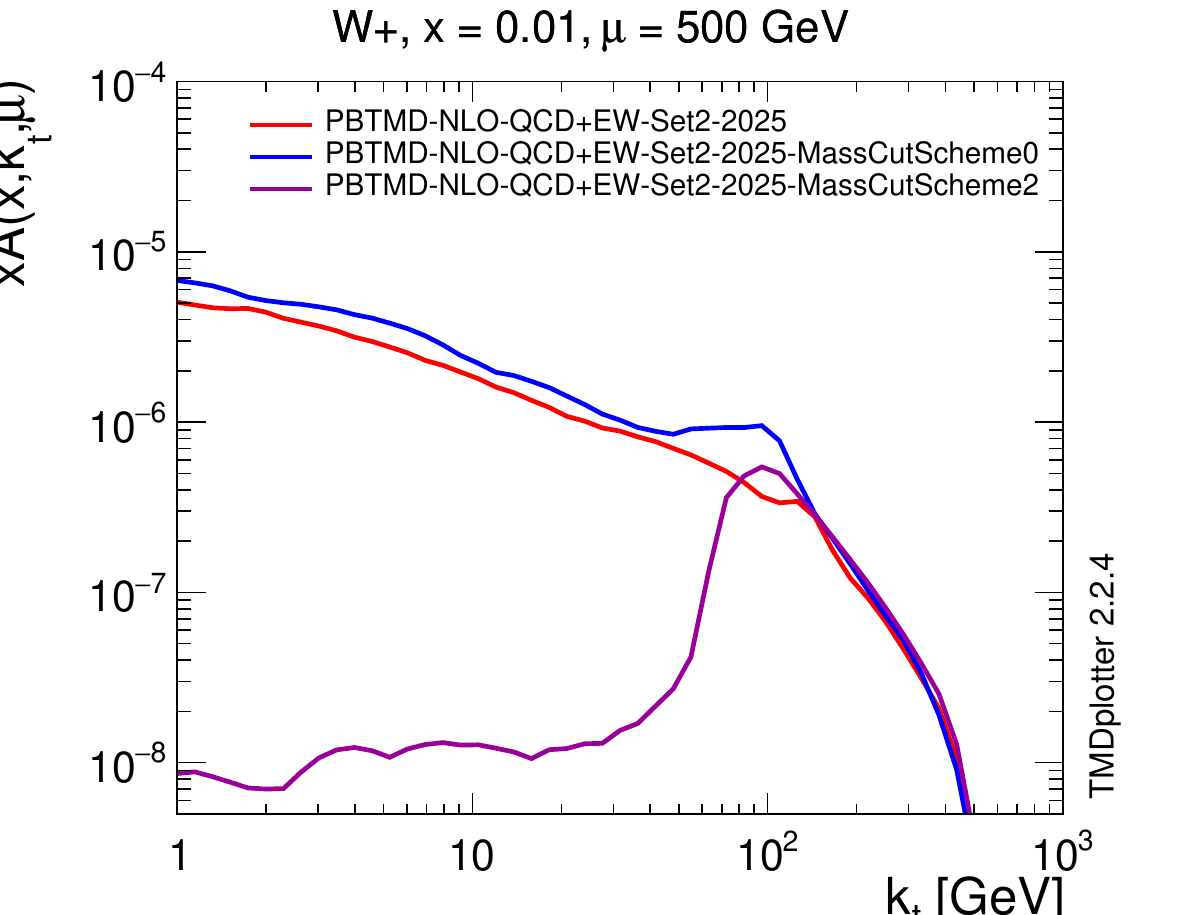}
\caption{\small The TMD density for the \PW -boson   for  \PBnewEW~Set2 at $\mu= 50$, 100  and $\mu= 500$ GeV as a function of $\kt$ for the different {\sc MassCutSchemes}.}
\label{EWFig3}
\end{center}
\end{figure}

We now turn to the discussion of the different  {\sc MassCutSchemes}: in Fig.~\ref{CollEWFig3} the collinear distribution for \PW -boson is shown as a function of $x$  for different $\mu$ as obtained with the different {\sc MassCutSchemes}.
The TMD distributions\footnote{In the TMD distributions fluctuations are visible which originate from the number of simulated events ($N=10^{10}$) in the forward evolution in \updfevolv .} allow further insights into the behavior of the different  {\sc MassCutSchemes}:
in Fig.~\ref{EWFig3} the TMD distribution for the \PW -boson is shown for different $\mu$ as a function of \kt\ as obtained with the different {\sc MassCutSchemes}.
Only {\sc MassCutScheme=1} gives a non-vanishing heavy boson density at low scales $\mu$, while in the other schemes there is a mass cut and even {\sc MassCutScheme=2 }(limiting the $z$-integral) results in a rather sharp threshold in \kt\ around the heavy boson mass (because of $z = 1 - M/q $). The contribution at small \kt\ obtained in {\sc MassCutScheme=2 }comes from multiple emissions which are added vectorially, indicating that at a scale $\mu=500$~\GeV\ the vector-boson is also undergoing evolution.
We observe that at $\mu=500$~\GeV\ the prediction from {\sc MassCutScheme=2} gives a larger density, compared to  the one with {\sc MassCutScheme=1}, because the suppression factor is still sizeable. 

\begin{tolerant}{9000}
For completeness,  the TMD distributions for heavy bosons with the default {\sc MassCutScheme=1} are shown in Fig.~\ref{EWFig2} of Appendix~\ref{TMDs}.
\end{tolerant}
In summary,  the densities of heavy bosons depend over a large range of scales $\mu$ on the treatment of the heavy boson mass: a dynamical suppression factor as used in {\sc MassCutScheme=1} offers a smooth description from small to high scales. The mass treatment  {\sc MassCutScheme=2}, motivated from angular ordering,  leads to modifications of the TMD, but also the collinear densities for scales $\mu < m_\PZ $. 
In order to obtain distributions applicable over a large range of $\mu$, {\sc MassCutScheme=1} is used as the default. 

\subsection{Validation of electroweak boson densities with DIS cross sections \label{validation}}

The DIS cross section is directly related to the collinear photon- and heavy boson parton densities. The derivation of the photon density from the DIS structure function $F_2$ has been discussed in Ref.\cite{Manohar:2017eqh}. Here we go a different path and use the measurement of the DIS cross section for validation of the obtained photon and heavy boson densities.

The single differential DIS cross section as a function of $Q^2$ for $e^+ p \to e^+ X$ with $\gamma$-exchange is given by:
\begin{eqnarray}
\frac{d \sigma_{NC}}{dQ^2} & = & \frac{2 \pi \alphaem^2} {Q^4 }  \int_{x_{\rm min}}^1 \frac{dx}{x} \left( 1 + (1-y)^2 \right) \nonumber \\ 
& & \times \left( e_u^2 \left[ xU(x,Q^2) + x{\bar U}(x,Q^2) \right] + e_d^2 \left[ xD(x,Q^2) + x{\bar D}(x,Q^2) \right]  \right)
\end{eqnarray}
with $Q^2,x,y$ being the DIS variables (for pure $\gamma$-exchange $xF_3=0$ ).
This equation can be written in terms of the photon density $f_\gamma$ of eq.~(\ref{gammaPDF}):
\begin{eqnarray}
\frac{d \sigma_{NC}}{dQ^2} & = & \frac{4 \pi^2 \alphaem} {Q^2 } \frac{ d xf_\gamma (x_{\rm min}, Q^2)}{d Q^2} 
\end{eqnarray}

The neutral current cross section coming from \PZ -exchange can be written, similarly, in terms of the \PZ -boson density $f_\PZ$:
\begin{eqnarray}
\frac{d \sigma_{NC}}{dQ^2} 
& = &  \frac{2 \pi \alphaem^2} { 16 \swfour \cwfour Q^4}  \left[\frac{Q^2}{(Q^2+M^2_\PZ) }\right]^2  (V_e^2 + A_e^2) \int \frac{dx}{x} \left( 1 + (1-y)^2 \right) \nonumber \\ 
& & \times \sum_{u,d} \left( (V_i^2 + A_i^2)  \left[ xF_i(x,Q^2) + x{\bar F_i}(x,Q^2) \right]  \right) \\ 
& = & \frac{\pi^2 \alphaem} {Q^2 }  \frac{V_e^2 + A_e^2}{ \swtwo \cwtwo}  \frac{ d xf_\PZ (x_{\rm min}, Q^2)}{d Q^2}  
\end{eqnarray}

Measurements of the DIS cross section~\cite{H1:2012qti}  can therefore be used to validate the boson densities obtained in the previous section, as the fit did not constrain the boson densities.

\begin{figure} [h!]
\centering
\includegraphics[width=0.45\linewidth]{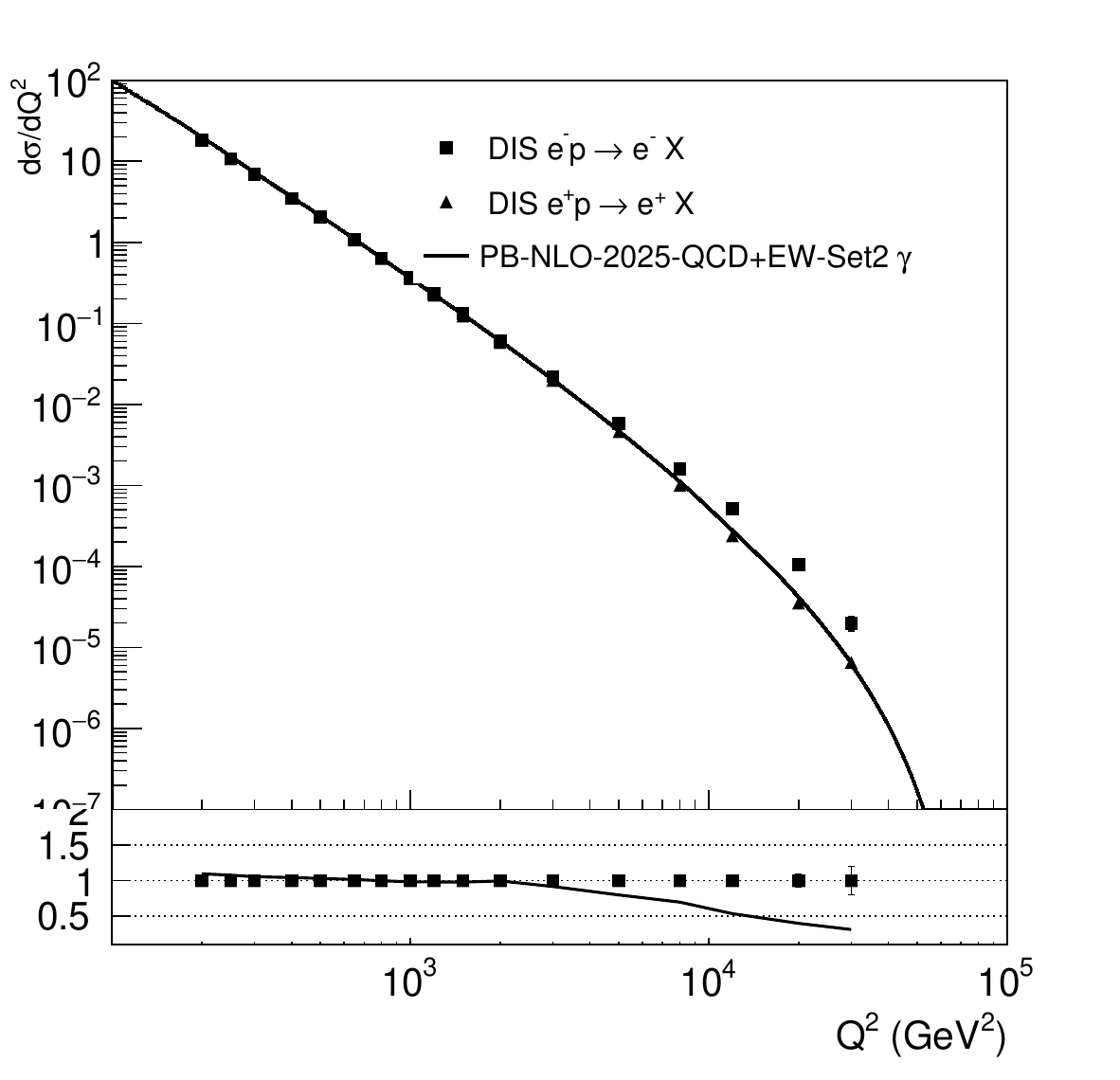} 
\includegraphics[width=0.45\linewidth]{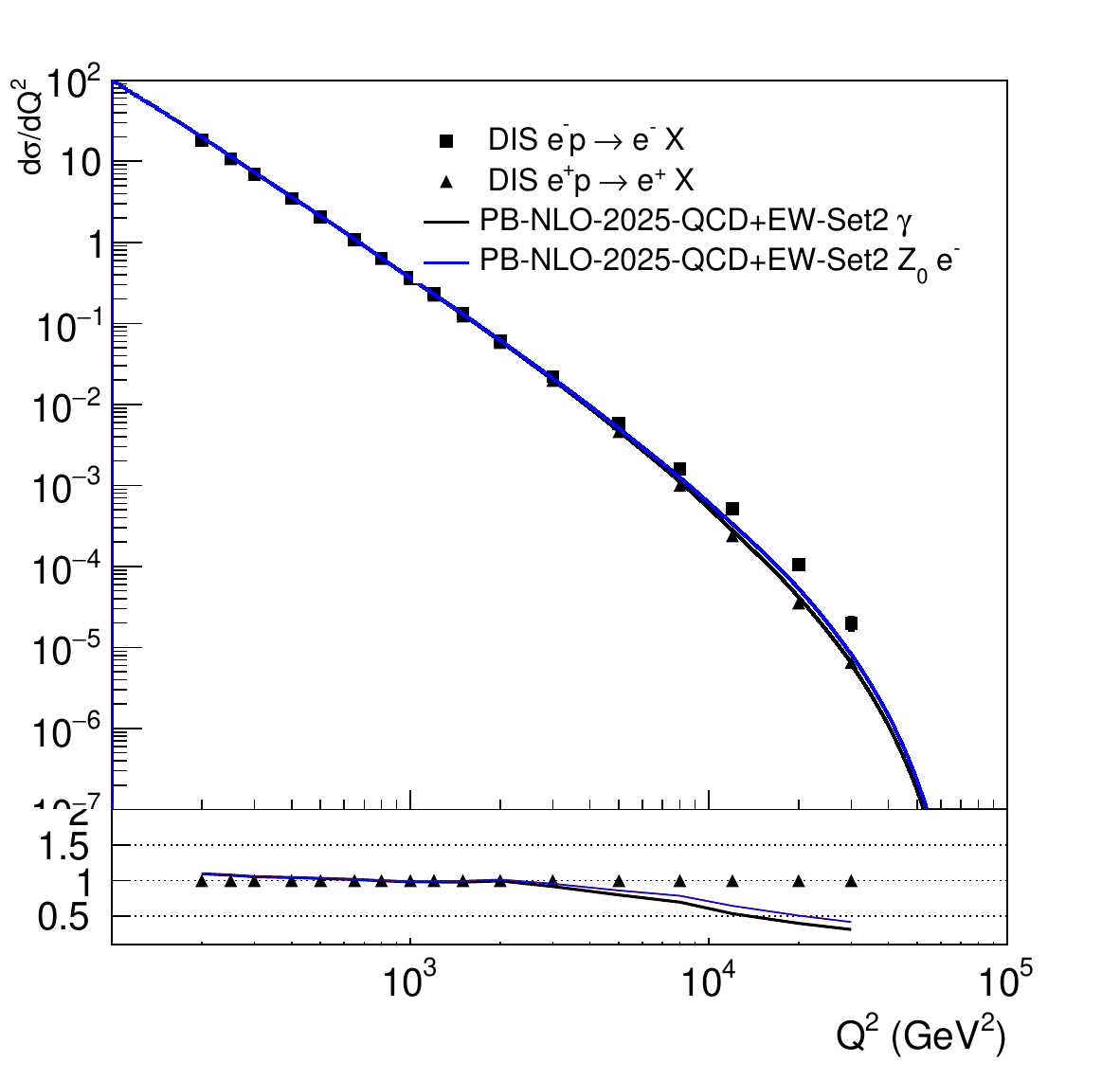} 
\caption {The measured DIS NC cross section \protect\cite{H1:2012qti}  compared to the calculation using the collinear photon density (left) and the sum of $\Pgamma + \PZ$ contribution (right). The ratio shows the prediction divided by the DIS $e^-p$ data set.
}
\label{fig:xsection_with_photon}
\end{figure}

A comparison of the measurement of the neutral current cross section $\frac{d \sigma_{NC}}{dQ^2} $ obtained in Ref.~\cite{H1:2012qti} with the calculation using photon and \PZ -boson densities is shown in Fig.~\ref{fig:xsection_with_photon}.  The DIS cross section is measured for $y_{\rm max} = 0.9$, which is translated into a lower limit on $x_{\rm min}=\frac{Q^2}{y_{\rm max} s }$. For the comparison we use the measurements with zero polarization $P_e=0$.
The predictions at low $Q^2$, where the photon exchange dominates, are in rather good agreement with the measurements. At larger $Q^2$, where also \PZ -exchange becomes important, the agreement becomes worse, since the interference terms included in $xF_3$ are not treated here. 
The inclusion of  \Pgamma -\PZ\ interference terms in the evolution will be addressed in future work.

The densities for \PW\, as determined in Sec.~\ref{EWdensities} cannot be directly used to describe the DIS cross section, since electrons carry electroweak quantum numbers and therefore they do not represent an inclusive summation over all initial-state weak isospin charges (see discussion in Ref.~\cite{Ruiz:2021tdt}), in contrast to quark states. Therefore, the different helicities must be included in the splitting functions (as detailed in eq.(2.7) of Ref.~\cite{Ruiz:2021tdt}). The DIS distributions  $f^{\rm DIS}_\PW$ are obtained using these splitting functions, instead of the unpolarized ones in eq.(\ref{W+PDF}).

\begin{figure} [h!]
\centering
\includegraphics[width=0.45\linewidth]{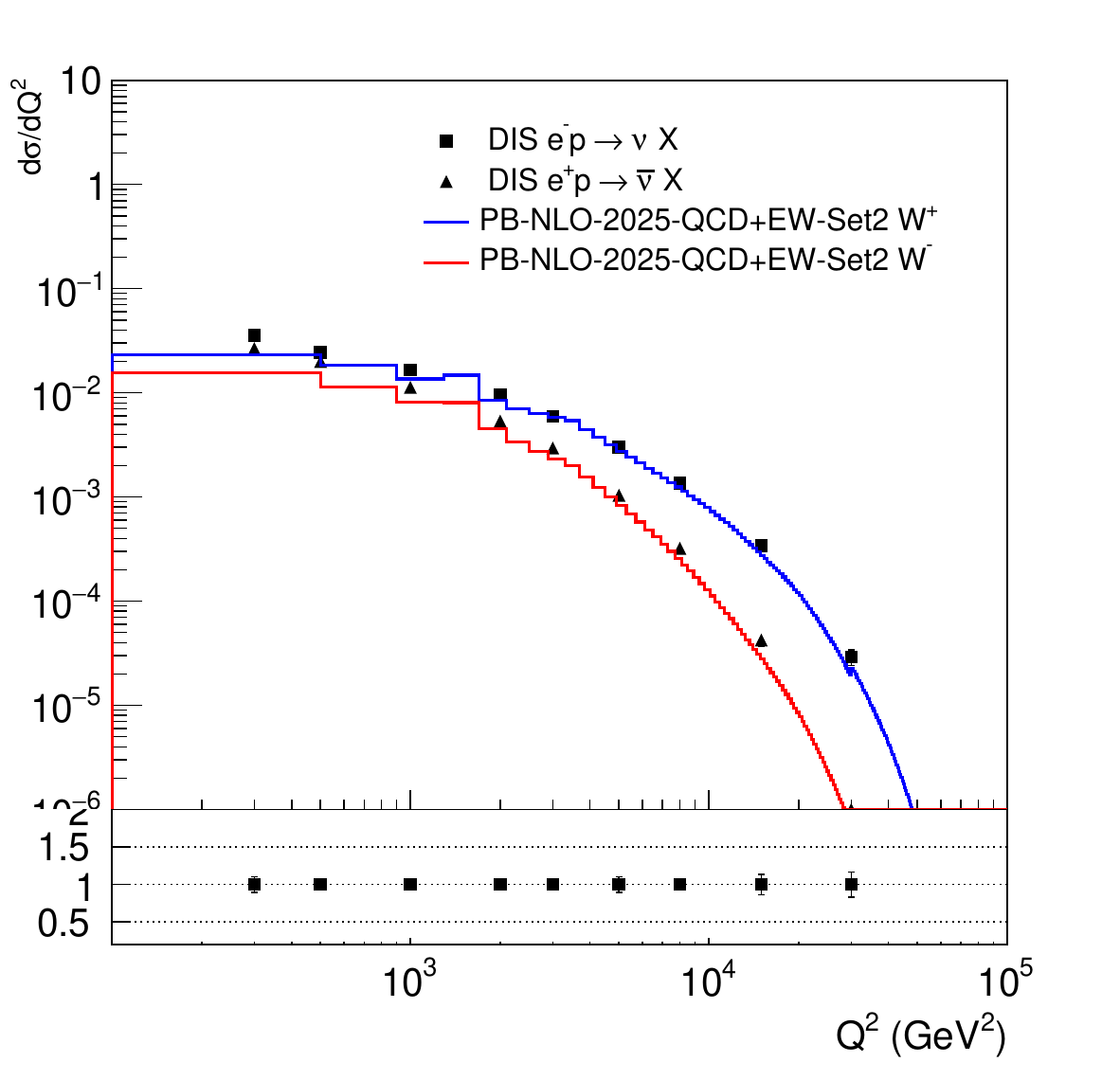} 
\caption {The measured DIS CC cross section \protect\cite{H1:2012qti}  compared to the calculation using the collinear \PW\ density. 
}
\label{fig:xsection_with_W}
\end{figure}

The charged-current DIS cross section for $e^+ p \to \nu X$ with $\PW^+$-exchange  can be written in terms of the the \PW -boson density $f_\PW$ (see eq.\ref{W+PDF}):
\begin{eqnarray}
\frac{d \sigma_{CC}}{dQ^2} & = & \frac{G_F^2}{2 \pi }  \left[\frac{M_\PW^2} {M^2_\PW + Q^2 }\right]^2  \int \frac{dx}{x}  \frac{1}{2} \left( 1 + (1-y)^2 \right) \left[ x{\bar U}(x,Q^2)  + x D (x,Q^2) \right] \\
& = & \frac{ 16 \pi \alphaem^2 }{128 \pi \swfour} \frac{1}{M_W^4} \left[\frac{M_\PW^2} {M^2_\PW + Q^2 }\right]^2  \int \frac{dx}{x}   \left( 1 + (1-y)^2 \right) \left[ x{\bar U}(x,Q^2)  + x D (x,Q^2) \right] \\
& = & \frac{ \pi^2 \alphaem }{Q^2} \frac{1}{ \swtwo} \frac{ d xf^{\rm DIS}_{\PW^+} (x_{\rm min}, Q^2)}{d Q^2} 
\end{eqnarray}
and similarly  with $\PW^-$-exchange (see eq.\ref{W-PDF}):
\begin{eqnarray}
\frac{d \sigma_{CC}}{dQ^2} & = &  \frac{ \pi^2 \alphaem }{Q^2} \frac{1}{ \swtwo}  \frac{d xf^{\rm DIS}_{\PW^-} (x_{\rm min}, Q^2) }{d Q^2}
\end{eqnarray}
In Fig.~\ref{fig:xsection_with_W} the measurement of the charged current DIS cross section is compared with predictions using the \PW -densities. The prediction follows the general trend of the measurements, especially at small $Q^2$ and the suppression at large $Q^2$. The predictions are within 20\% for the $\PW^-$ case. 

\section{Conclusion\label{Conclusion}}

We presented a unified determination of QCD and electroweak parton densities using an extended DGLAP evolution, which is solved iteratively with the Parton Branching method allowing for a treatment of kinematic relations at each branching vertex.  We apply collinear splitting functions for QCD partons at next-to-leading order accuracy, while for the photon and heavy bosons we use leading order splitting functions and couplings, in the so-called "phenomenological" scheme. Interference of \Pgamma\  and \PZ\ during the evolution is not considered.

The initial distributions for quarks and gluons are fitted to DIS precision data recorded at HERA. 
Including photon and heavy-boson contributions does not significantly alter the fit parameters or the overall fit quality. The Parton Branching method allows the simultaneous determination of transverse momentum dependent densities for the photon and heavy bosons as well. 

The photon and heavy boson densities are applied to calculate the neutral and charged current cross section in deep inelastic scattering as a function of $Q^2$ and the comparison with measurements from HERA showed reasonable agreement. This comparison favors the treatment of the heavy boson mass as an additional propagator term, in contrast to an approach with a sharp mass cut or those which limit the $z$-integration by a mass dependent term.

The collinear and TMD densities for QCD partons as well as for photons and heavy electroweak bosons are, to our knowledge, the first such determinations directly fitted to DIS precision data.
The densities are available in LHAPDF and TMDlib formats.

\vskip 0.5 cm 
\begin{tolerant}{8000}
\noindent 
{\bf Acknowledgements.} 
We are grateful for many discussions with A.~Bagdatova, S.~Baranov,  A.~Kotikov, A.~Lipatov, M.~Malyshev G.~Lykasov and the other participants of the WeeklyOffline Meeting during the past years. 
We thank A. Glazov for constructive comments and suggestions.
\end{tolerant}

\appendix

\section*{Appendices}

\section{Input parameters and input scheme \label{EWinput}}
We work in the $(\alpha_0, G_\mu, m_\PZ )$ scheme~\cite{Chiesa:2024qzd}, with input parameters $\alpha_0$, $G_\mu$ and the mass of the \PZ -boson $m_\PZ$.  The advantage of this scheme is that the independent parameters are well known experimentally with high precision.
In this scheme, the weak mixing angle $\swtwo$ as well as the mass of the \PW -boson $m_\PW $ are derived from the input parameters, at lowest order given by:
\begin{equation}
(1- \swtwo) \swtwo = \frac{\pi \alpha_0} {\sqrt{2 G_\mu m_\PZ^2}}
\end{equation}
$$ \cwtwo = 1 -\swtwo ,  \;\; \cwtwo = \frac{m^2_\PW}{m^2_\PZ}$$  
\begin{equation}
\swtwo = \frac{1}{2} - \sqrt{\frac{1}{4} - \frac{\pi \alpha_0} {\sqrt{2} G_\mu m^2_\PZ}}, \;\; m^2_\PW = \frac{\pi\alpha_0}{\sqrt{2} G_\mu \swtwo}
\end{equation}

We list below the EW input parameters~\cite{ParticleDataGroup:2024cfk}:

$$ G_\mu = 1.166 378 8(6)  \times 10^{-5}\, \GeV^{-2}$$ $$ \alpha_0 = 1/137.035 999 084(21) $$ 
$$M_\PZ = 91.1880(20)\, \GeV$$  $$M_\PW = 80.3692(133)  \, \GeV$$

\section{Electroweak splitting functions \label{EWsplitt}}

In Tab.~\ref{Tab:GenSplitt}  we give an overview of the splitting functions at leading order. A full account including different polarization states is given in Refs.~\cite{Dittmaier:2025htf,Ciafaloni:2024alq,Ciafaloni:2005fm,Fornal:2018znf,Chen:2016wkt,Bauer:2017isx}.
\begin{table}[h]
\begin{tabular}{ l l l l  l l }
\label{Tab:GenSplitt}
Splitting function & Diagram & $V=\Pgluon$  & $V=\Pgamma$  & $V= \PZ$  & $V= \PW$  \\  \hline  \\
 $P_{qq}(z) = \frac{\alphaeff}{2 \pi} \frac{1+z^2}{1-z}$ & 
\begin{minipage}[t]{0.09\textwidth}
\vspace*{-0.3cm} 
\rotatebox{0.}{\scalebox{0.38}{\includegraphics{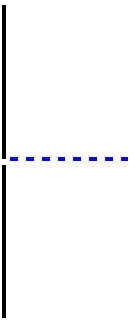}}}
\end{minipage}
& $ C_F  \as $
& $ \alpha_{em}$ 
& $\frac{\alpha_{em} (V_f^2 + A_f^2)  }{4 \sin^2 \theta_W \cos^2 \theta_W }  $
&  $\frac{\alpha_{em}  |V_{qq}|^2}{4 \sin^2 \theta_W  }$
 \\ \\ \hline \\
$P_{qV}(z)= \frac{\alphaeff}{2 \pi} \left(z^2 + (1-z)^2\right)$ & 
\begin{minipage}[t]{0.09\textwidth}
\vspace*{-0.3cm} 
\rotatebox{0.}{\scalebox{0.38}{\includegraphics{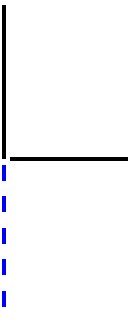}}}
\end{minipage}
& $ T_R \as $
& $ \alpha_{em}$  
& $\frac{\alpha_{em} (V_f^2 + A_f^2)}{4 \sin^2 \theta_W \cos^2 \theta_W } $
& $ \frac{\alpha_{em}  |V_{qq}|^2}{4 \sin^2 \theta_W  }$
\\ \\  \hline \\ 
$P_{Vq}(z) = \frac{\alphaeff}{2 \pi} \frac{1+(1-z)^2}{z}$ & 
\begin{minipage}[t]{0.09\textwidth}
\vspace*{-0.3cm} 
\rotatebox{0.}{\scalebox{0.38}{\includegraphics{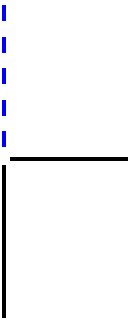}}}
\end{minipage}
& $C_F  \as $ 
&  $\alpha_{em}$ 
& $\frac{\alpha_{em} (V_f^2 + A_f^2)}{4 \sin^2 \theta_W \cos^2 \theta_W } $
& $ \frac{\alpha_{em}  |V_{qq}|^2}{4 \sin^2 \theta_W  }$
\\ \\ \hline \\
$P_{VV'} (z) = \frac{\alphaeff}{2 \pi} 2 \left( \frac{1-z}{z} + \frac{z}{1-z} + z(1-z)\right)$ & 
\begin{minipage}[t]{0.09\textwidth}
\vspace*{-0.3cm} 
\rotatebox{0.}{\scalebox{0.38}{\includegraphics{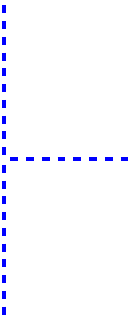}}}
\end{minipage}
& $ C_A  \as $
&  $0$   
& \multicolumn{2}{ c}{
$4 \pi \alpha_{em} \cot^2 \theta_W$}
\\ \\  \hline
\end{tabular}
\caption{Splitting functions at leading order and $\alphaeff$ for QCD and electroweak processes with the color factors $C_F=4/3$, $C_A=3$ and $T_R=1/2$. }
\end{table}

\section{Transverse Momentum Distributions \label{TMDs}}
\begin{figure}[h!tb]
\begin{center} 
\includegraphics[width=0.32\textwidth,angle=0]{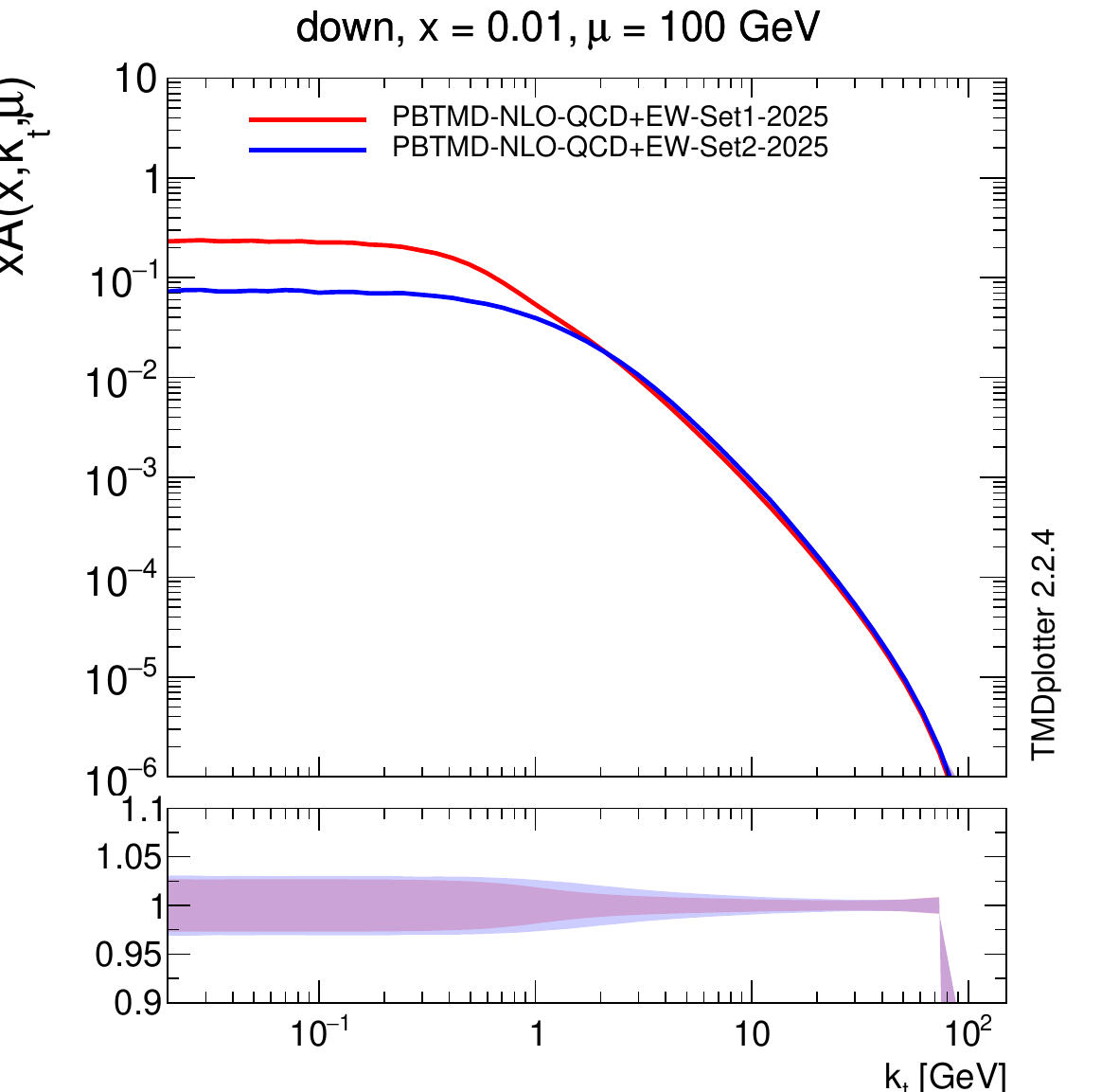}
\includegraphics[width=0.32\textwidth,angle=0]{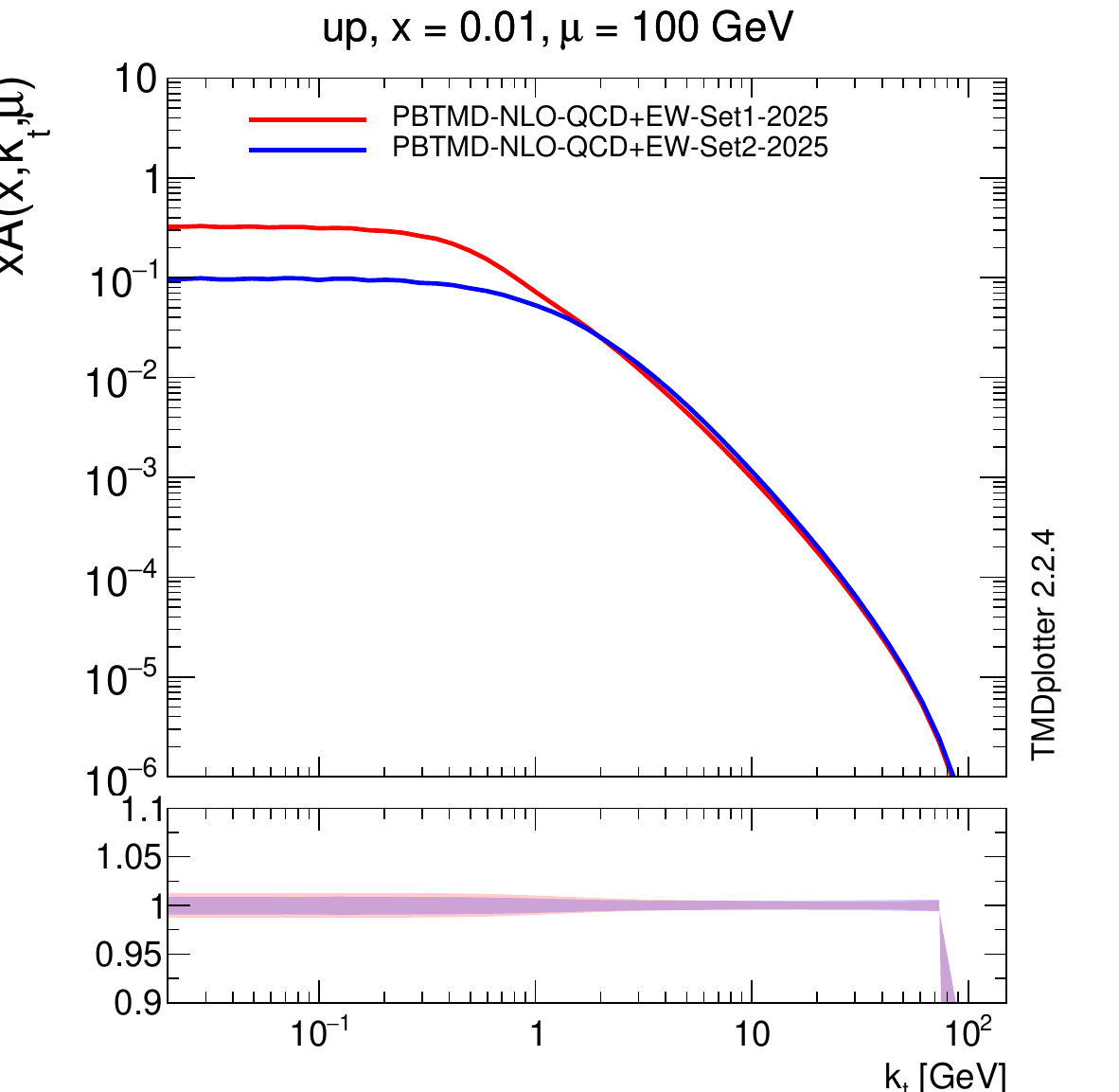}
\includegraphics[width=0.32\textwidth,angle=0]{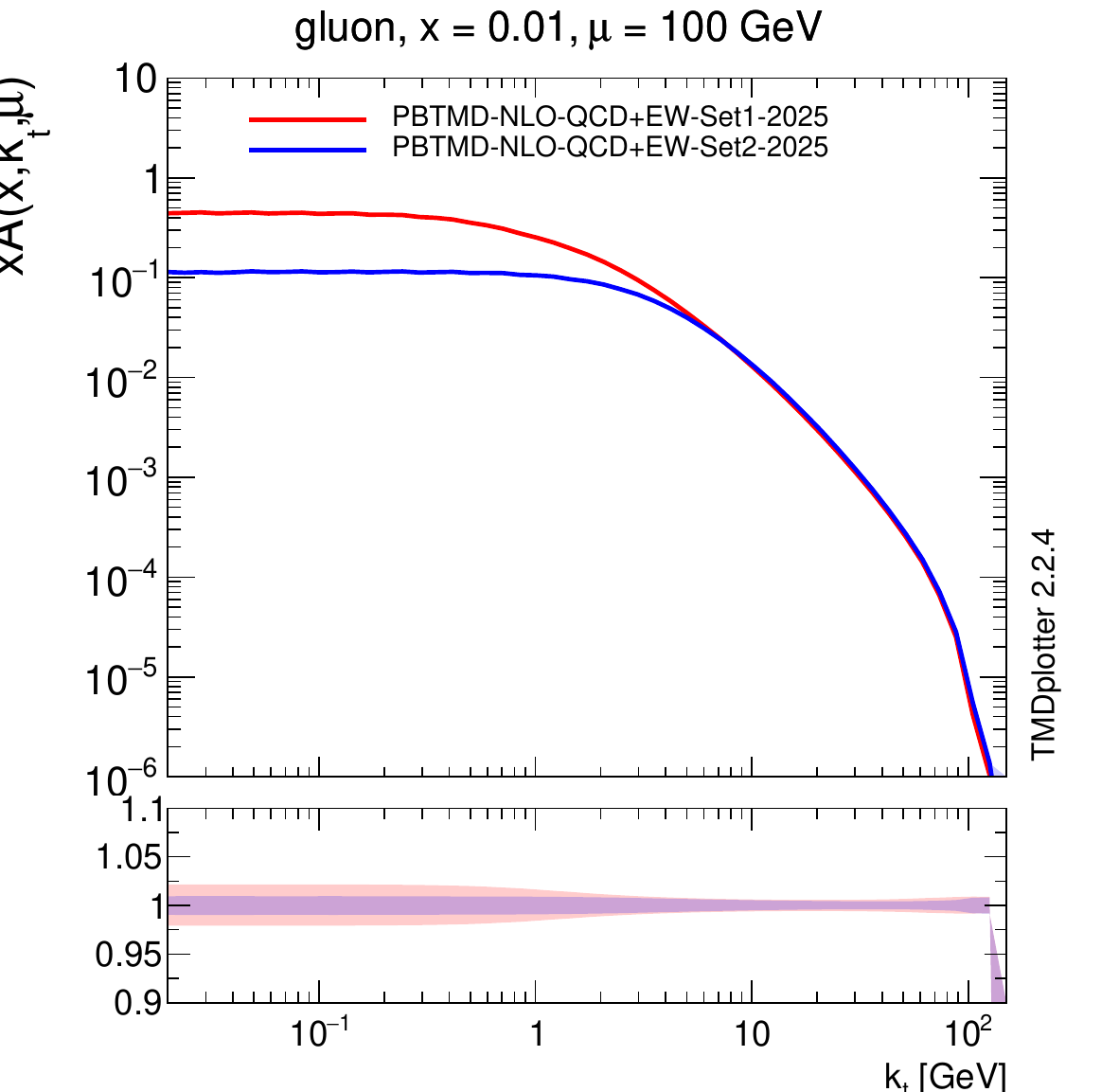}
\caption{\small The transverse momentum dependent quark and gluon densities at $\mu= 100$ GeV as a function of \kt . Shown are \PBnewEW~Set1 and Set2. The uncertainty band includes only uncertainties coming from the measurements, the model uncertainties are not shown explicitly (see text). }
\label{TMDFit-Set1+2}
\end{center}
\end{figure}

\begin{figure}[h!tb]
\begin{center} 
\includegraphics[width=0.32\textwidth,angle=0]{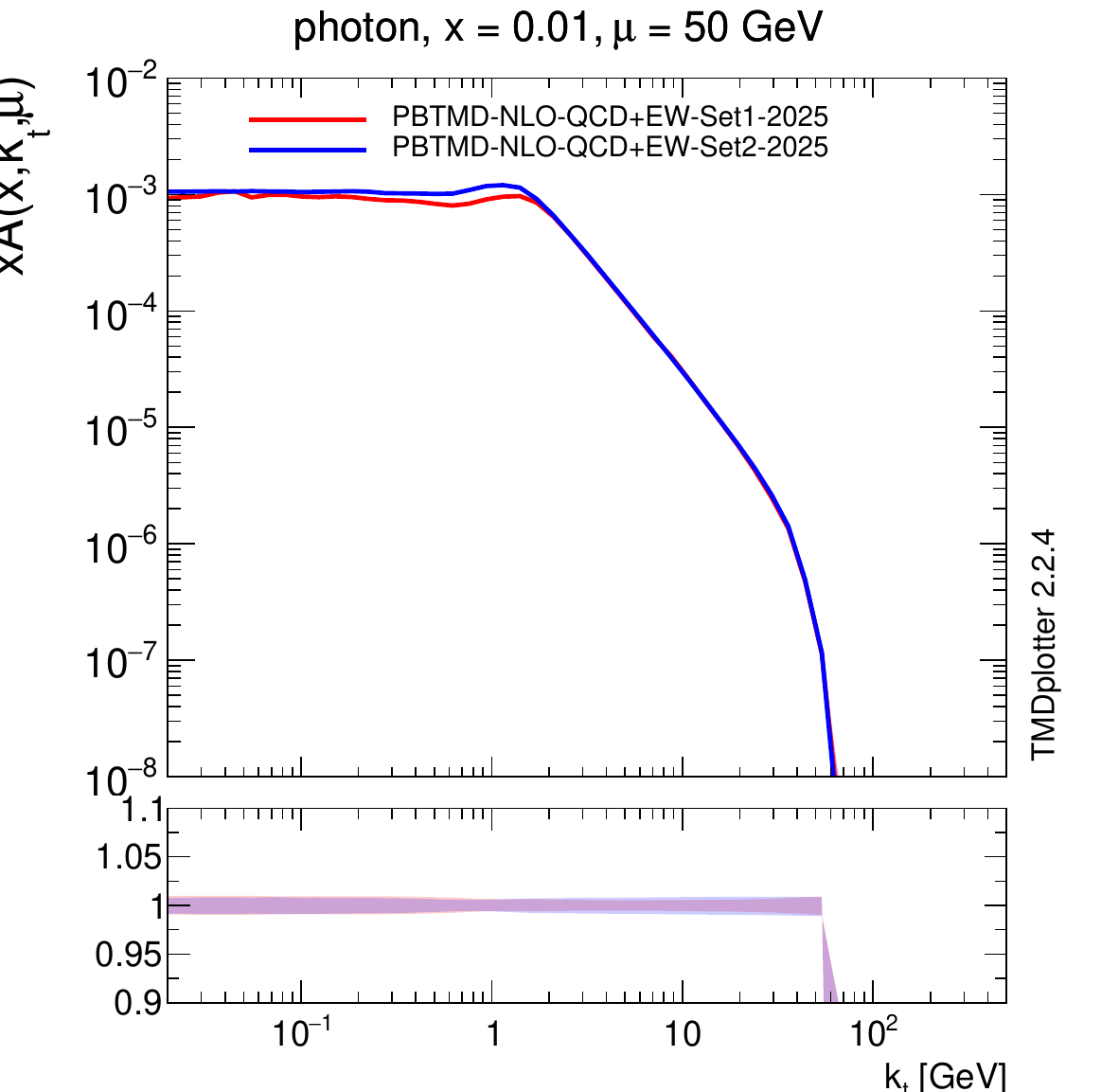}
\includegraphics[width=0.32\textwidth,angle=0]{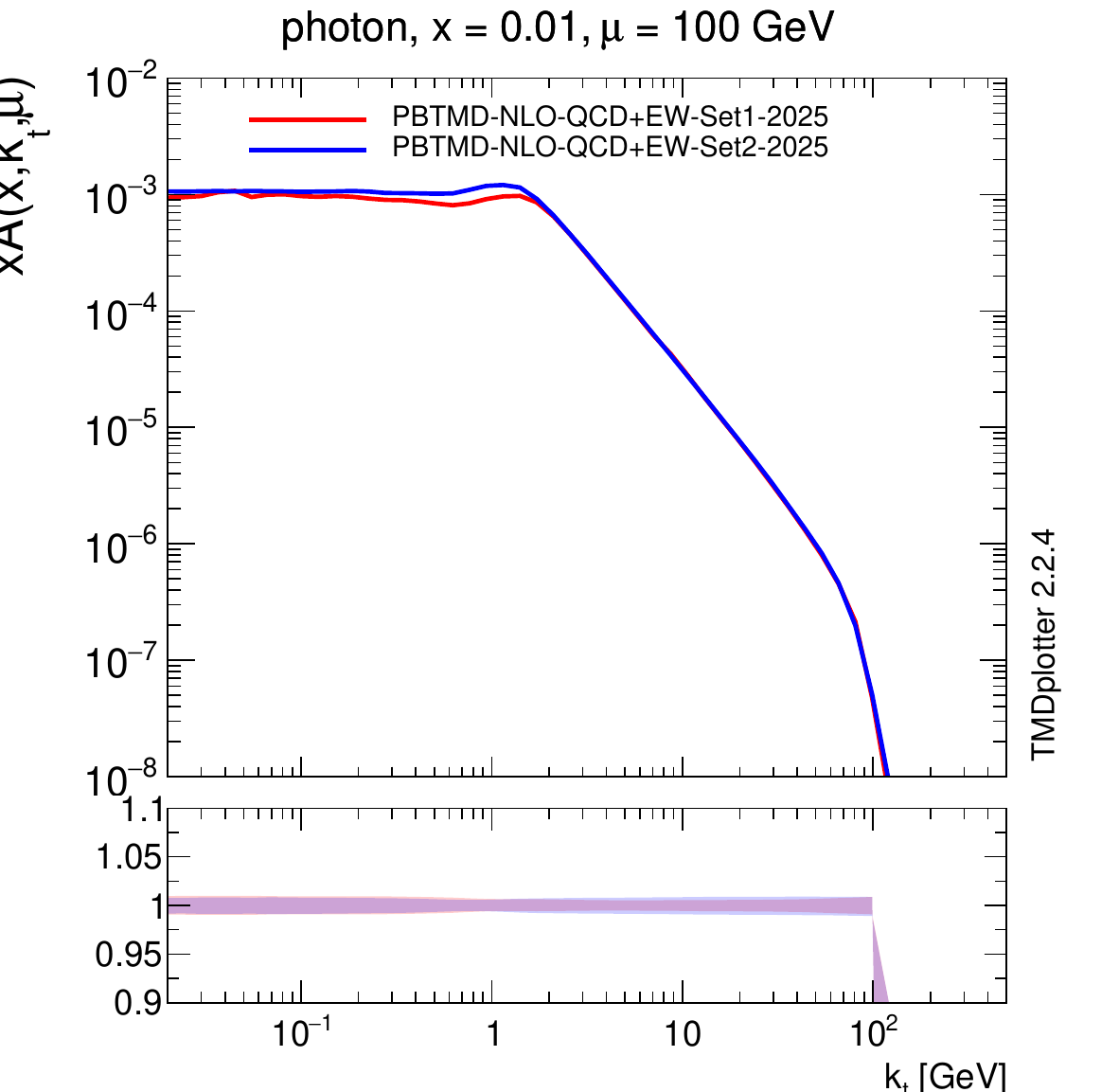}
\includegraphics[width=0.32\textwidth,angle=0]{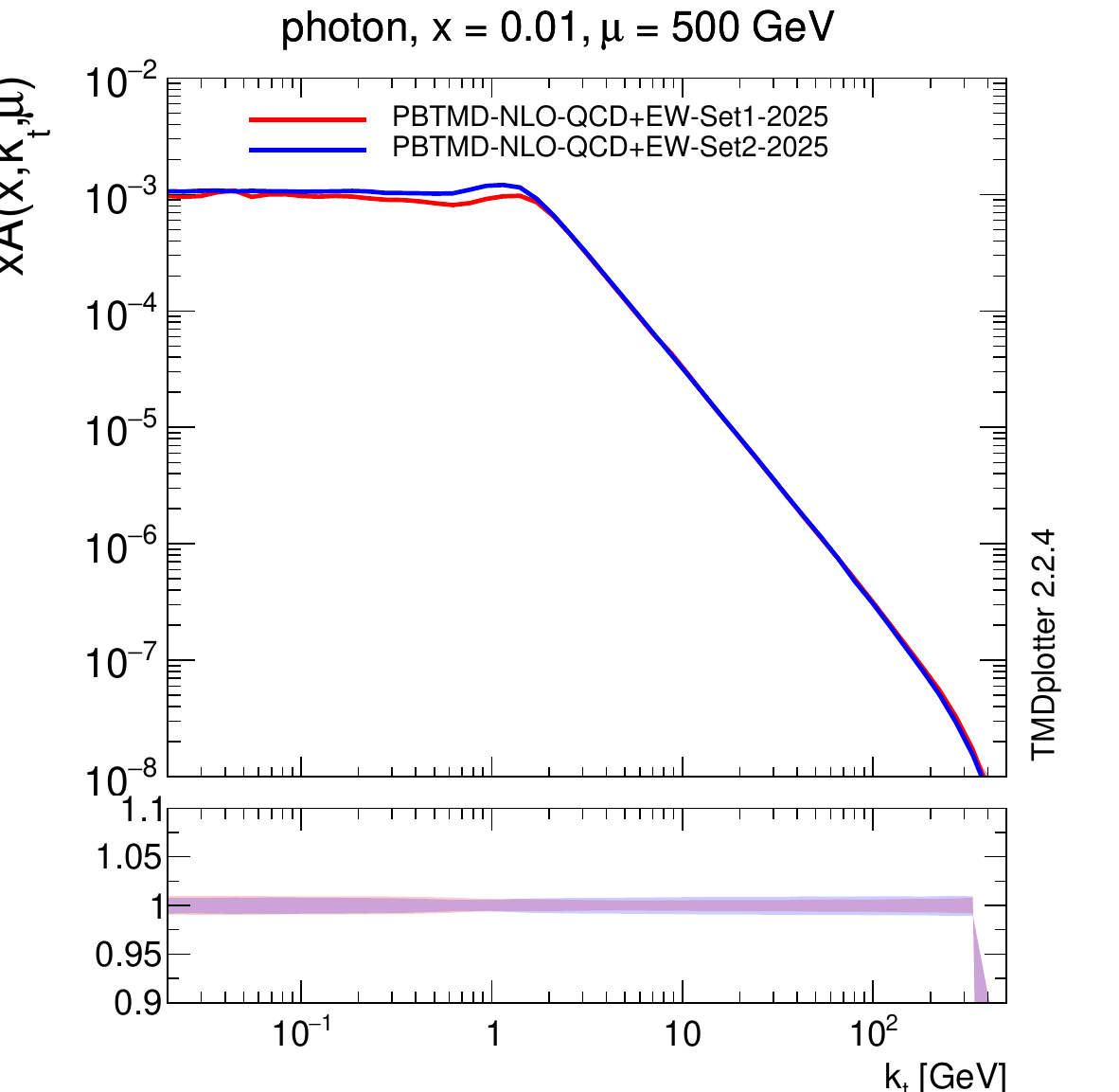}
\caption{\small The TMD photon density for  \PBnewEW~Set1 and Set2 at $\mu= 50$, 100 and 500~GeV  as a function of $\kt$  including uncertainties coming from the measurements, the model uncertainties are not shown explicitly (see text).  }
\label{PhotonFig2}
\end{center}
\end{figure}

\begin{figure}[h!tb]
\begin{center} 
\includegraphics[width=0.32\textwidth,angle=0]{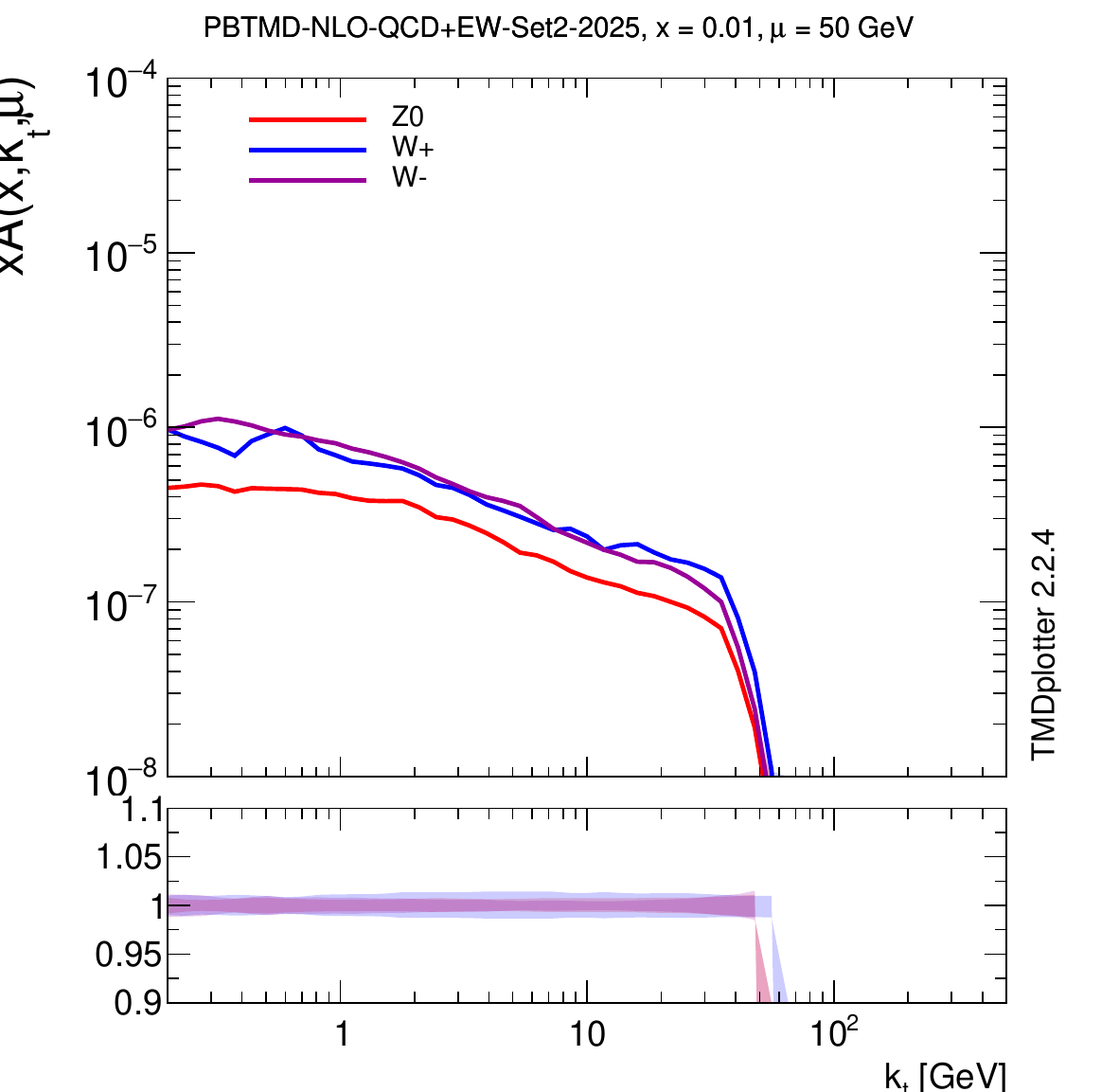}
\includegraphics[width=0.32\textwidth,angle=0]{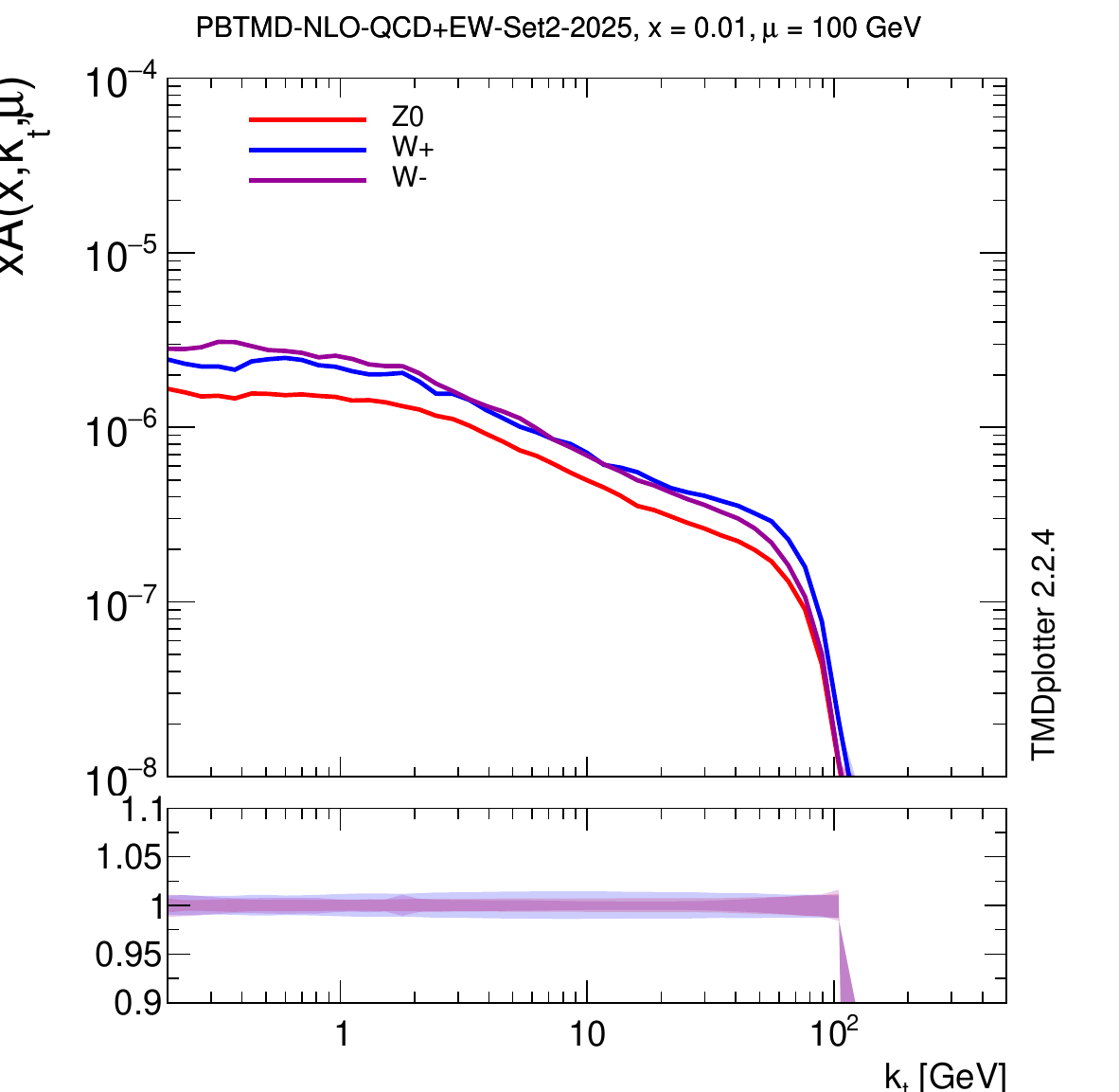}
\includegraphics[width=0.32\textwidth,angle=0]{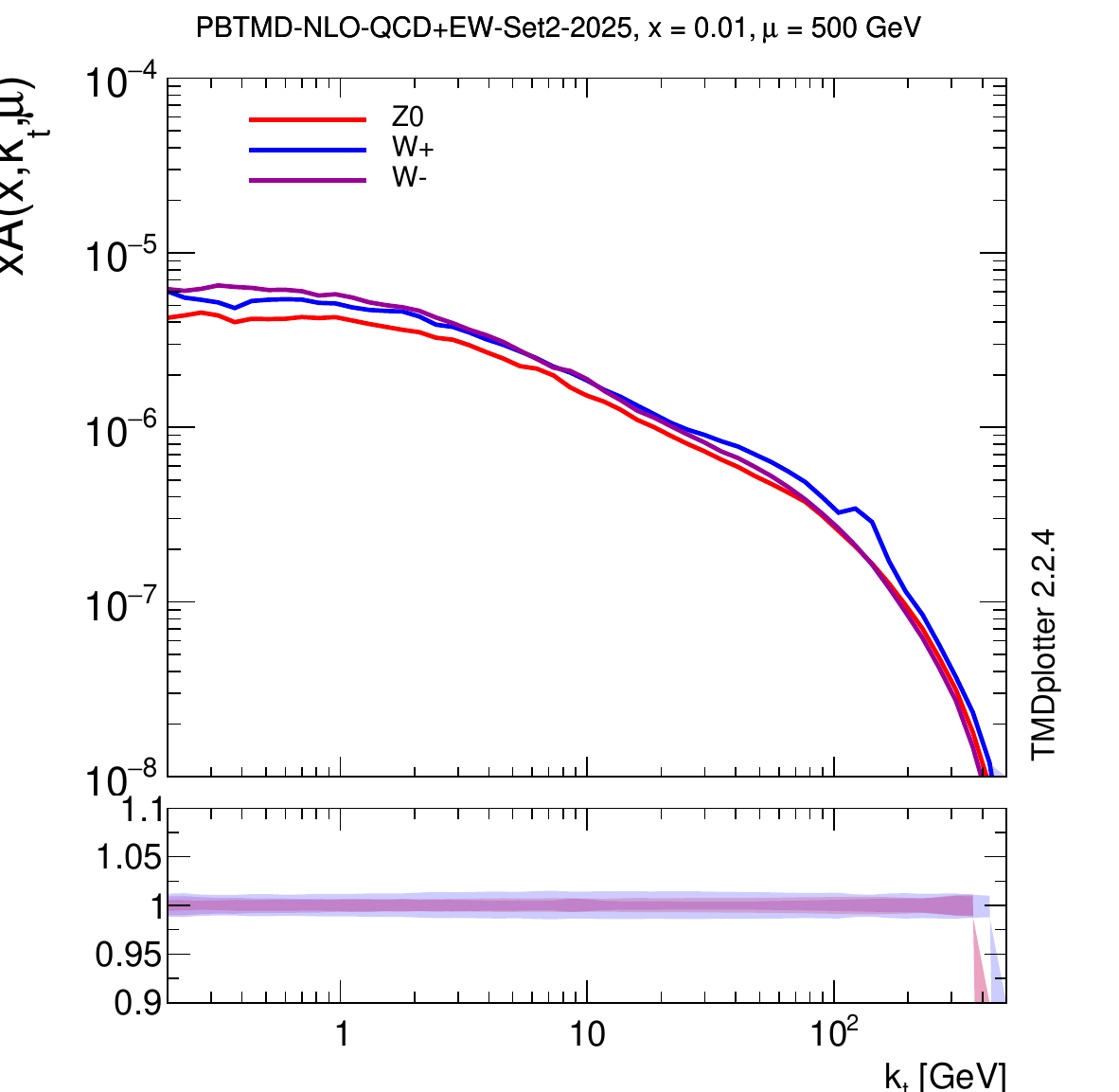}
\caption{\small The TMD vector-boson density for  \PBnewEW~Set2 at $\mu= 50$ GeV and $\mu= 100$ GeV as a function of $\kt$  including uncertainties coming from the measurements, the model uncertainties are not shown explicitly (see text).}
\label{EWFig2}
\end{center}
\end{figure}

In Fig.~\ref{TMDFit-Set1+2} we show the transverse momentum distribution of quarks and gluons  at $\mu= 100$ GeV and $x=0.01$, including only uncertainties coming from the measurements, the model uncertainties are not shown explicitly. Especially for the bosons in Fig.~\ref{PhotonFig2} and~\ref{EWFig2}, one can observe fluctuations in the distributions, which come from the number of simulated events ($N=10^{10}$) in \updfevolv , applying a Monte Carlos approach with forward evolution. This approach suffers from large weight fluctuations, especially for the case of heavy bosons, resulting in fluctuations  in the differential \kt -distributions. The model uncertainties, as obtained from the integrated distributions (see Figs.~\ref{CollEWSet1} and \ref{CollEWSet2}) are of the same order as the statistical fluctuations obtained in the TMD distributions, and are therefore not shown.

In TMD density for photons is shown in Fig.~\ref{PhotonFig2}, 
the one for heavy bosons in Fig.~\ref{EWFig2}.

\bibliographystyle{mybibstyle-new.bst}
\raggedright  
\providecommand{\href}[2]{#2}\begingroup\raggedright\endgroup

\end{document}